\documentclass[reprint,preprintnumbers,
amsmath,amssymb,aps,nofootinbib,superscriptaddress]{revtex4-1}
\usepackage[pdftex]{graphicx}
\usepackage{dcolumn}
\usepackage{subfigure}
\usepackage{bm}
\usepackage{natbib}
\usepackage{cases}
\usepackage{hyperref}
\hypersetup{
    colorlinks=true,
    linkcolor=blue,
    citecolor=blue,
    filecolor=blue,
    urlcolor=blue,
    }
\makeatletter \def\@eqnnum{{\normalsize \normalcolor (\theequation)}}  \makeatother %

\begin{document}

\preprint{\begin{minipage}[b]{1\linewidth}
\begin{flushright} KEK-TH-1890 \\ KEK-Cosmo-189 \end{flushright}
\end{minipage}}

\title{Higgs vacuum metastability in primordial inflation, preheating, and reheating}
\author{Kazunori Kohri}
\email{kohri@post.kek.jp}
\affiliation{KEK Theory Center, IPNS, KEK, Tsukuba, Ibaraki 305-0801, Japan}
\affiliation{The Graduate University of Advanced Studies (Sokendai),Tsukuba, Ibaraki 305-0801, Japan}
\author{Hiroki Matsui}
\email{matshiro@post.kek.jp}
\affiliation{The Graduate University of Advanced Studies (Sokendai),Tsukuba, Ibaraki 305-0801, Japan}

\begin{abstract}
  Current measurements of the Higgs boson mass and top Yukawa coupling
  suggest that the effective Higgs potential develops an instability
  below the Planck scale. If the energy scale of inflation is as high
  as the GUT scale, inflationary quantum fluctuations of the Higgs field
  can easily destabilize the standard electroweak vacuum and produce a
  lot of AdS domains.  This destabilization during inflation can be
  avoided if a relatively large nonminimal Higgs-gravity or
  inflaton-Higgs coupling is introduced.  Such couplings
  generate a large effective mass term for the Higgs, which can raise
  the effective Higgs potential and suppress the vacuum fluctuation of
  the Higgs field. After primordial inflation, however, such effective
  masses drops rapidly and the nonminimal Higgs-gravity or
  inflaton-Higgs coupling can cause large fluctuations of the Higgs field to be generated via parametric
  resonance, thus producing AdS domains in the preheating
  stage. Furthermore, thermal fluctuations of the Higgs field cannot be
  neglected in the proceeding reheating epoch.  We discuss the Higgs
  vacuum fluctuations during inflation, preheating, and reheating, and
  show that the Higgs metastability problem is severe unless the energy scale of  
  the inflaton potential is much lower than the GUT scale.
\end{abstract}
\maketitle

\section{Introduction}
With the discovery of the Higgs boson at the LHC, the standard
model has been completed, and elementary particle physics has entered
a new era. The recent measurements of the Higgs boson mass,
$m_{h}=125.09\ \pm \  0.21\ ({\rm stat})\ \pm \  0.11\ ({\rm syst})\ {\rm GeV}$
~\cite{Aad:2015zhl,Aad:2013wqa,Chatrchyan:2013mxa,Giardino:2013bma}
and top quark mass,
$m_{t}=173.34\pm 0.27\ ({\rm stat})\pm 0.71\ ({\rm syst})\ {\rm GeV}$~\cite{2014arXiv1403.4427A}
suggest that the running of the quartic Higgs self-coupling $\lambda$
becomes negative, and the effective Higgs potential becomes unstable at the
scale $\Lambda_{I} = 10^{10} \sim 10^{11} \ {\rm GeV}$~\cite{Buttazzo:2013uya}.
\\

If the effective Higgs potential is unstable below the Planck scale, our
electroweak vacuum is metastable and should eventually decay into the true vacuum through quantum tunneling~\cite{Kobzarev:1974cp,Coleman:1977py,Callan:1977pt}. The timescale for this decay, however, is longer than the age of the Universe, so it was thought that the Higgs vacuum metastability does not
phenomenologically have any significant impact on the observed
Universe~\cite{Degrassi:2012ry,Isidori:2001bm,Ellis:2009tp,EliasMiro:2011aa}.
However, recently it has been argued that the electroweak vacuum instability during inflation or at the end of inflation might threaten the existence of the
Universe~\cite{Espinosa:2007qp,Fairbairn:2014zia,Kobakhidze:2013tn,Lebedev:2012sy,Kobakhidze:2013tn,Enqvist:2013kaa,Herranen:2014cua,Kobakhidze:2014xda,Kamada:2014ufa,Enqvist:2014bua,Hook:2014uia,Kearney:2015vba,Espinosa:2015qea, Gross:2015bea, Herranen:2015ima, Ema:2016kpf}.  Stochastic quantum fluctuations produced during
inflation can cause the Higgs field value to grow as  
\begin{equation}
\left< { h }^{ 2 } \right> \simeq \frac { { H }^{ 3 }t }{ 4{ \pi  }^{ 2 } },
\end{equation}
where $h$ is the value of the Higgs and $H$ is the Hubble expansion rate
(or the Hubble scale).  If the Higgs field evolves beyond the
instability scale $\Lambda_{I}$ before the end of inflation, the Higgs field
classically rolls down into the true vacuum and Anti-de Sitter (AdS)
domains are formed, which is potentially catastrophic.

Not all AdS domains generated during inflation threaten the existence of our
Universe~\cite{Hook:2014uia,Kearney:2015vba,Espinosa:2015qea}, with the
significance highly depending on the number and the evolution of the AdS
domains.  In Ref.\cite{Espinosa:2015qea}, the authors discussed how
AdS domains evolve both during inflation and after the end of inflation. The
Higgs AdS domains can either shrink or expand, eating other regions of the
electroweak vacuum. Although high-energy-scale inflation can lead to the generation of
more expanding AdS domains during inflation, such domains never take
over all of the inflationary dS space, because the inflationary expansion
always overcomes the expansion of the AdS domains. However, after the
inflationary epoch, although some AdS domains harmlessly shrink, others expand 
and devour our whole Universe. This indicates that the existence of AdS domains 
in our observable universe is catastrophic, and so in this paper we focus on 
the conditions for them not to be generated.

The generation of Higgs AdS domains during inflation can be suppressed
by introducing a relatively large nonminimal Higgs-gravity or
inflaton-Higgs coupling.  Such couplings give rise to large inflationary
effective mass terms, which raise the effective Higgs potential and
weaken the Higgs vacuum
fluctuations~\cite{Espinosa:2015qea,Lebedev:2012sy,Kamada:2014ufa}.
However, at the end of the inflation, such mass terms drops rapidly, 
and become ineffective for stabilizing the Higgs field~\cite{Kamada:2014ufa}.
Nonminimal Higgs-gravity or inflaton-Higgs coupling can also cause large Higgs
fluctuations to be generated via parametric resonance~\cite{Herranen:2015ima, Ema:2016kpf},
thus producing a lot of AdS domains in the preheating stage.  
Moreover, thermal Higgs fluctuations are
not negligible in the reheating epoch after inflation.  
In this paper, we analyse the vacuum fluctuations of the
Higgs field during inflation, preheating, and reheating and show
that the Higgs metastability is a serious problem in the inflationary
Universe unless the energy scale of the inflaton potential is much lower than
the GUT scale or the effective Higgs potential is stabilized below the
Planck scale. In this paper, we use the reduced Planck mass, $M_{\rm pl}=2.4\times10^{18}\ {\rm GeV}$.

%%%%%%%%%%%%%%%%%%%%%%%%%%%%%%%%%%%%%%%%%%%%%%%%%%%%%%%%%%%%%%%%%%%%%%
\section{Inflationary Higgs fluctuations and Higgs AdS domains}
%%%%%%%%%%%%%%%%%%%%%%%%%%%%%%%%%%%%%%%%%%%%%%%%%%%%%%%%%%%%%%%%%%%%%%
In this section, we discuss the evolution of the massless Higgs field
during inflation using the Fokker-Planck equation, and determine the
probability that Higgs AdS domains are formed. In the large-field regime
$h \gg v$, where $v=246\ {\rm GeV}$, 
the effective Higgs potential can
be approximated by the following Renormalization Group (RG) improved
tree-level expression,\footnote{The effective Higgs potential is not
gauge-invariant, but physical quantities extracted from the effective
potential ( Higgs boson mass, S-matrix elements, tunneling rates) are
gauge-invariant~\cite{DiLuzio:2014bua,Andreassen:2014eha,Andreassen:2014gha}.
However, here we ignore the gauge dependence of the effective Higgs
potential for simplicity.}
\begin{equation}
V_{\rm eff}\left( h \right)=\frac{\lambda_{\rm eff}\left( h \right)}{4}h^{4},
\end{equation}
where $\lambda_{\rm eff}\left( h \right)$ is the effective self-coupling
including the RG improved couplings, the one-loop corrections.
The instability scale $\Lambda_{I}$
can be defined as the effective self-coupling $\lambda_{\rm eff}\left( h \right)$
becomes negative at the scale.
In the RG improved effective Higgs potential, the instability scale is $\Lambda_{I}\simeq h_{\rm max}$
where $h_{\rm max}$ defined as $V_{\rm eff}\left( h \right)$ takes its maximal value.\footnote{If the effective Higgs potential has large effective mass terms $m_{\rm eff}^{2}h^{2}/2$ 
or includes one-loop thermal correction $\Delta { V }_{\rm eff}\left( h,T \right)$
at the high temperature, the instability scale doesn't coincide with $h_{\rm max}$, i.e. $h_{\rm max}\gtrsim \Lambda_{I}$,
and we cannot assume $h_{\rm max}=\Lambda_{I}$. }

When the Hubble scale $H$ is smaller than the Higgs field value at
the maximum of the effective Higgs potential, $h_{\rm max}$, 
the Higgs field can tunnel into the true vacuum via the Coleman-de Luccia
instanton~\cite{Coleman:1980aw}. If the Hubble rate is as
large as the maximal Higgs field value $h_{\rm max}$,
the transition is dominated by the Hawking-Moss instanton~\cite{Hawking:1981fz}. 
The transition probability of the Higgs field during inflation can also be
obtained by statistical approaches using the Fokker-Planck equation,
with the result being approximately equal to that obtained using the
Hawking-Moss instanton~\cite{Linde:1991sk}.

The Fokker-Planck equation describes the evolution of the probability
$P\left(h,t\right)$ that the Higgs takes the value $h$ in one Hubble horizon-size region at cosmic time
$t$, and takes the following form
\begin{equation}
\frac { \partial { P } }{ \partial t } =\frac { \partial^{2}  }{ \partial h^{2} } \left(\frac { { H }^{ 3 } }{ 8{ \pi  }^{ 2 } }P\right)+ \frac { \partial }{ \partial h} \left(\frac { V_{\rm eff}'\left( h \right) }{ 3H } { P }\right),
\end{equation}
where the prime $'$ denotes the derivative with respect to the field, i.e.
$V'_{\rm eff}(h) = \frac{dV_{\rm eff}}{dh}$. 
According to Ref.\cite{Espinosa:2015qea}, we may ignore the gradient of the effective Higgs potential
$V_{\rm eff}'\left( h \right)$,\footnote{This assumption is reasonable for $H^2 \gtrsim 0.01  V_{\rm eff}''$.} 
with assuming that the field value is $h=0$ at $t=0$, which gives,
\begin{equation}
P\left( h,t \right) =\sqrt { \frac { 2{ \pi  } }{ { H }^{ 3 }t }  } \exp\left( -\frac { 2{ { \pi  }^{ 2 } } }{ { H }^{ 3 }t } { h }^{ 2 } \right).
\label{eq:hjfg}
\end{equation}
During inflation, if the Higgs in some region evolves to values larger than $\Lambda_{I}$, 
then it will roll into the true vacuum and a potentially dangerous AdS region will be formed. 
The survival probability of the electroweak vacuum at the end of inflation is
estimated to be~\cite{Espinosa:2007qp,Espinosa:2015qea}
\begin{eqnarray}
{ P }\left(  h<\Lambda_{I},N_{\rm tot} \right) &\equiv& \int _{ -\Lambda_{I} }^{ \Lambda_{I} }{ dh\ P\left( h,t_{\rm end}\right)  },\label{eq:hsdsg} \\ &=& {\rm erf}\left( \frac { \sqrt {2}\pi\Lambda_{I} }{ H\sqrt { N_{\rm tot} }  }  \right),\label{erf0}
\end{eqnarray}
where $t_{\rm end}$ denotes the time at the end of inflation, $N_{\rm
tot}$ is the total $e$-folding number, defined as $N_{\rm tot}=H\cdot
t_{\rm end}$, and ${\rm erf}\left( x \right)$ is the error function,
which for $x\gg 1$ is approximately given by
\begin{equation}\label{erf1}
{\rm erf}\left( x \right) \simeq 1-\frac { 1 }{ \sqrt { \pi } x } { e }^{ -{ x }^{ 2 } }.
\end{equation}
Note that the total e-folding number can be much larger than the
observable e-folding number $N_{\rm hor}$, which is the number of
e-foldings before the end of inflation that the largest observable
scales left the Horizon, i.e. we could have ${N }_{ \rm tot }\gg{ N }_{
\rm hor }$.\footnote{The $e$-folding number that corresponds to when the
current horizon scale left the horizon is almost the same as that
associated with large-scale CMB observations, and we have ${ N }_{\rm
hor }\simeq { N }_{ \rm CMB }\simeq 60$} This implies that our
observable universe is only part of the whole Universe.

$P(h,t)$ in the Fokker-Planck equation describes the probability
distribution of the Higgs in one horizon-sized
region~\cite{Nambu:1989uf,Nambu:1988je}. Inflation produces many such
regions, and our observable Universe contains $e^{3N_{\rm hor}}$ of
them. As such, the survival probability can be estimated as
\begin{equation}
\left\{ { P }\left(  h<\Lambda_{I} ,N_{\rm tot} \right) \right\}^{{ e }^{3N_{\rm hor}}}  > \frac{1}{2},
\end{equation}
which, using \eqref{erf0} and \eqref{erf1}, can be approximately re-written as
\begin{equation}
\left\{ 1-\frac { H\sqrt { N_{\rm tot} }  }{ \pi \sqrt { 2\pi  }\ \Lambda_{I} } { e }^{ -{ \frac { 2{ \pi  }^{ 2 }\Lambda_{I}^{ 2 } }{ { H }^{ 2 }N_{\rm tot}}  } }\right\}^{{ e }^{3N_{\rm hor}}}  > \frac{1}{2}.
\label{eq:ssfdfd}
\end{equation}
If we set $N_{\rm hor}=60$ and $N_{\rm tot}=10^{3}$
in~(\ref{eq:ssfdfd}),  we obtain the following upper bound on the Hubble scale,
\begin{equation}
\frac{ H }{\Lambda_{I}}<1.1\times10^{-2}.
\end{equation}
%%

%%%%%%%%%%%%%%%%%%%%%%%%%%%%%%%%%%%%%%%%%%%%%%%%%%%%%%%%%%%%%%%%%%%%%%
\begin{figure}[t]
\includegraphics[width=87mm]{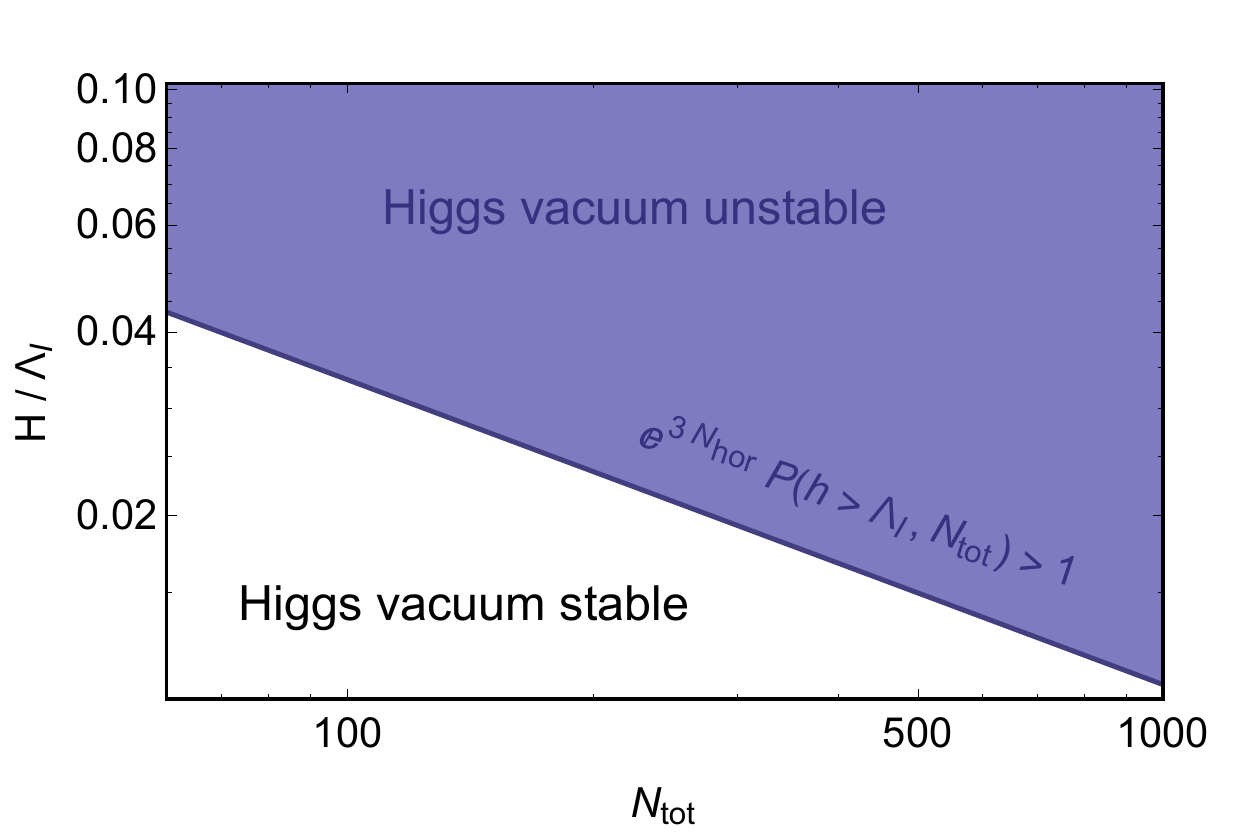}
\caption{Plot of the upper bound on $H/\Lambda_{I}$ as a function of
  the total $e$-folding number $N_{\rm tot}$, as determined by
  (\ref{eq:horsdsd}). We set $N_{\rm hor}=60$ and
  $60 \le N_{\rm tot}$. }
\label{Fig:1}
\end{figure}
%%%%%%%%%%%%%%%%%%%%%%%%%%%%%%%%%%%%%%%%%%%%%%%%%%%%%%%%%%%%%%%%%%%%%%

Alternatively, we can restrict $H$ by using the probability that the field rolls down into the true vacuum at the end of inflation, ${ P }\left( h>\Lambda_{I},N_{\rm tot}\right)$, which is given as
\begin{eqnarray} { P }\left( h>\Lambda_{I},N_{\rm tot}\right)
  &=& 1- { P }\left( h<\Lambda_{I},N_{\rm tot} \right), \\
  &\simeq& \frac { H\sqrt { N_{\rm tot} } }{ \pi \sqrt { 2\pi }\ \Lambda_{I}
   } { e }^{ -{ \frac { 2{ \pi }^{ 2 }\Lambda_{I}^{ 2 } }{ { H }^{ 2 }N_{\rm tot}} } }.
\end{eqnarray}
Multiplying this by ${ e }^{3N_{\rm hor}}$ gives the number of AdS domains in
the local region corresponding to our observable Universe.  As such, the condition that there be no AdS regions within the current horizon is expressed as
\begin{equation}
{ e }^{3N_{\rm hor}}{ P }\left(  h>\Lambda_{I},N_{\rm tot} \right) < 1,
\label{eq:horsdsd}
\end{equation}
which can be approximated as
\begin{equation}
  \scalebox{0.9}{$\displaystyle { e }^{3N_{\rm hor}}{ P }\left(  h>\Lambda_{I},N_{\rm tot}\right) =
    \frac { H\sqrt { N_{\rm tot } }  }{ \pi \sqrt { 2\pi  }\ \Lambda_{I} } { e }^{ 3N_{\rm hor}-{ \frac { 2{ \pi  }^{ 2 }\Lambda_{I}^{ 2 } }{ { H }^{ 2 }N_{\rm tot} }}}$}.\label{eq:lllfdf}
\end{equation}
Therefore, we have the upper bound on the Hubble scale,
\begin{equation}
\frac { H }{ \Lambda_{I} } <\sqrt { \frac { 2{ \pi  }^{ 2 } }{ 3N_{\rm hor}N_{\rm tot}}  }.
\label{eq:hkkksdg}
\end{equation}
If we take $N_{\rm hor}=60 $ and $N_{\rm tot}=10^{3}$ in
(\ref{eq:lllfdf}), the upper bound on the Hubble scale is numerically found to be
\begin{equation}
\frac{ H }{\Lambda_{I}}<1.1\times10^{-2}.
\end{equation}

In Fig.~\ref{Fig:1}, we plot the upper bound on $H/\Lambda_{I}$ as a
function of the total $e$-folding number $N_{\rm tot}$, as determined by (\ref{eq:horsdsd}). We assume that $N_{\rm hor}=60$ and that the total $e$-folding number is greater than $N_{\rm hor}$, i.e. $60 \le N_{\rm tot}$. It is natural to consider that the total $e$-folding number $N_{\rm tot}$ may be huge, due to the stochastic
nature of inflation,  and the inflationary Higgs vacuum
fluctuations grow as time goes by.  Consequently, requiring the Higgs vacuum to remain stable throughout inflation puts tight constraints on the Hubble scale during inflation.

Although the AdS domains impact on the existence of our observable
universe, the expansion of AdS domains never takes over the expansion
of inflationary dS space~\cite{Espinosa:2015qea}, and therefore, it is
impossible that one AdS domain terminates the inflation on all the
space of the Universe. However, If the proportion of non-inflating
domains or the AdS domains dominates all the space of the
Universe~\cite{Kearney:2015vba,Sekino:2010vc}, the inflating space
would crack, and inflation comes to an end.

%%%%%%%%%%%%%%%%%%%%%%%%%%%%%%%%%%%%%%%%%%%%%%%%%%%%%%%%%%%%%%%%%%%%%%
\section{Inflationary Higgs vacuum fluctuations during inflation}
%%%%%%%%%%%%%%%%%%%%%%%%%%%%%%%%%%%%%%%%%%%%%%%%%%%%%%%%%%%%%%%%%%%%%%

In the previous section we discussed the massless Higgs vacuum
fluctuations during inflation, and by solving the Fokker-Planck equation
we were able to determine the probability for the formation of Higgs AdS
domains. In general, the inflationary Higgs fluctuations become as large
as the Hubble scale $H$ during inflation. However, if the Higgs field
has a large effective mass, the Higgs vacuum fluctuations are suppressed
during inflation. Field fluctuations in the massive case, particularly
in the case where $m>3H/2$, have often been discussed using different
descriptions in the literature.  In this section we introduce mass
terms for the Higgs field, determine its fluctuations, calculate the
probability for the formation of Higgs AdS domains and obtain
constraints on the model parameters by requiring consistency with
observations.

%%%%%%%%%%%%%%%%%%%%%%%%%%%%%%%%%%%%%%%%%%%%%%%%%%%%%%%%%%%%%%%%%%%%%%
\subsection{Fluctuations of light Higgs field}
%%%%%%%%%%%%%%%%%%%%%%%%%%%%%%%%%%%%%%%%%%%%%%%%%%%%%%%%%%%%%%%%%%%%%%

The FLRW metric  is given by
\begin{equation}
g_{\mu\nu}={\rm diag}\left( -1,\frac { { a }^{ 2 }\left( t \right)  }{ 1-K{ r }^{ 2 } } ,{ a }^{ 2 }\left( t \right) { r }^{ 2 },{ a }^{ 2 }\left( t \right) { r }^{ 2 }\sin^{2} { \theta  }  \right) ,
\end{equation}
where $K$ is the curvature constant and $a = a(t)$ is the scale factor.
For simplicity we will take $K=0$.  Then, the scalar curvature is
obtained as
\begin{equation}
R=6\left[ 
{ \left( \frac { \dot { a }  }{ a }  \right)  }^{ 2 }+\left( \frac { \ddot { a }  }{ a }  \right)  
\right]. 
\end{equation}
In a de Sitter Universe where $a\propto e^{Ht}$, the Ricci scalar is
estimated to be $R\simeq 12H^{2}$.  We assume that the total scalar potential for the
inflaton and Higgs is given as follows
\begin{equation}
V\left( \phi,h \right)=V_{\rm inf}\left( \phi \right)+V_{\rm eff}\left( h \right)+\frac{1}{2}\xi h^{2} R
+\frac{1}{2}g^{2}\phi^{2}h^{2},
\end{equation}
where $\phi$ is the inflation field, $\xi$ is the nonminimal Higgs-gravity
coupling constant, and $g$ is the coupling constant
between $h$ and $\phi$.  The Klein-Gordon equation for Fourier modes 
of the Higgs field is given as 
\begin{equation}
{ \delta \ddot { h }  }_{ k }+3H{ \delta \dot { h }  }_{ k }+\left( \frac{k^{2} }{a^{2}}+\xi R+g^{2}\phi^{2}    \right){ \delta { h }  }_{ k }=0,\label{eq:kdfd}
\end{equation}
where we have assumed that we can neglect the contribution from 
$V_{\rm eff}(h)$ in comparison with the other terms.
A finite value of $\xi$ or $g$ therefore generates
an effective Higgs mass,
which during inflation is approximately given as
\begin{equation} 
 m_{\rm eff}^{2}\simeq 12 H^{2}\xi+g^{2}\phi^{2}.
\end{equation}
The additional Higgs mass can raise the effective Higgs potential and
suppress the vacuum fluctuation of the Higgs field.  
The maximum of the Higgs potential gets shifted to larger values of $h$.

We introduce the redefined field $\delta\sigma_{ k
}$ which is related to $\delta h_k$ as
\begin{equation}
 \delta \sigma_{ k }=a\delta { h }_{ k }.
\end{equation}
The Klein-Gordon equation for $\delta \sigma_{ k }$ takes the form
\begin{equation}
 \delta \sigma_{ k }''+\left(k^{2}-\frac{1}{\tau^{2}}\left(\nu^{2}-\frac{1}{4}\right)\right) \delta \sigma_{ k }=0\label{eq:kjkk},
\end{equation}
where the conformal time has been introduced and is defined as $d\tau =
dt/a$, and $\nu$ is defined to be
\begin{equation}
\nu=\sqrt { \frac{9}{4}-\frac{m_{\rm eff}^{2}}{H^{2}} }.
\end{equation}
%%
%%%%%%%%%%%%%%%%%%%%%%%%%%%%%%%%%%%%%%%%%%%%%%%%%%%%%%%%%%%%%%%%%%%%%%
\begin{figure*}[t]
        \begin{tabular}{cc}
	\begin{minipage}{0.5\hsize}
		\centering
		\subfigure[\ nonminimal coupling $\xi$ ]{
		\includegraphics[width=87mm]{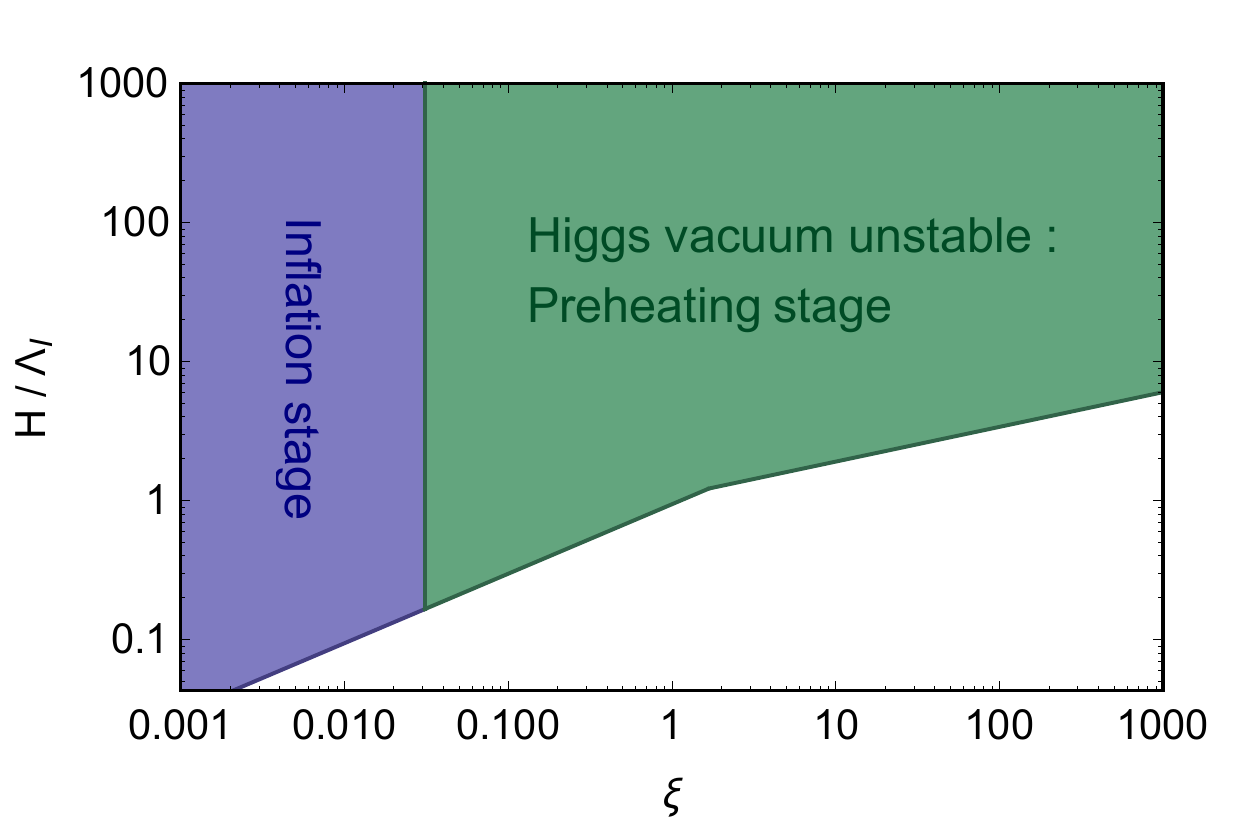}
		\label{Fig:2a}}\end{minipage}
\begin{minipage}{0.5\hsize}
		\centering
		\subfigure[\ nonminimal coupling $\xi$ ]{
		\includegraphics[width=87mm]{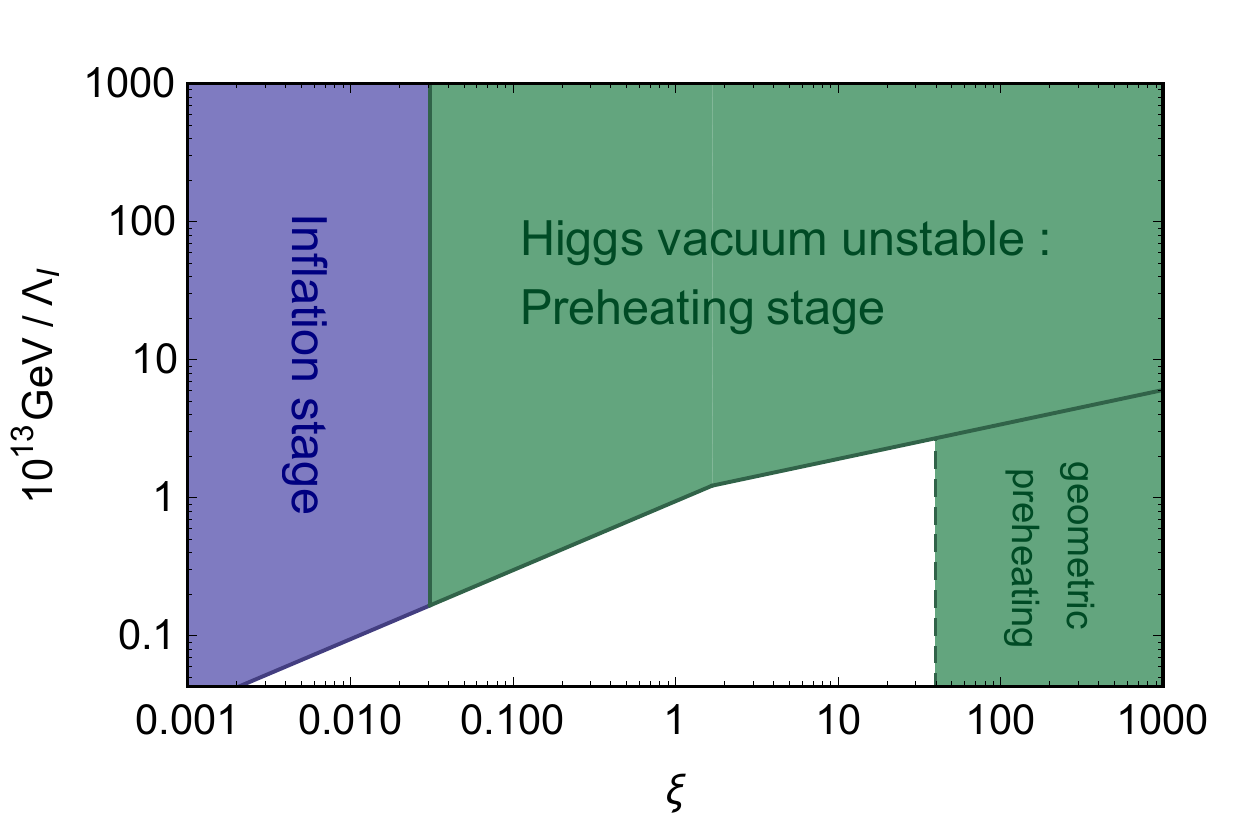}
		\label{Fig:2b}}
\end{minipage}
\end{tabular}
\caption{Constraints on $H/\Lambda_{I}$ as a function of the
  nonminimal coupling $\xi$.  We have plotted the lines shown in
  (\ref{eq:nndfdf}) and (\ref{eq:kbdfdfbb}). We took the effective mass to be
  $m_{\rm eff}=\sqrt{12\xi}H$ and set $N_{\rm hor}=N_{\rm tot}=60$.
  In the left panel we neglect the constraint coming from the parametric amplification of the
  Higgs field in order to remove model-dependence. On the other hand, in
  the right panel we include the constraint coming from broad resonance
  during the preheating stage, taking the quadratic chaotic
  inflation model $V_{\rm inf}\left( \phi \right) 
  \simeq \frac12m_{\phi}^{2}\phi^{2}$ and the Hubble scale $H\simeq 10^{13}\ {\rm  GeV}$ as an example.  }\label{Fig:2}
\end{figure*}
%%%%%%%%%%%%%%%%%%%%%%%%%%%%%%%%%%%%%%%%%%%%%%%%%%%%%%%%%%%%%%%%%%%%%%
The general solution of Eq.~(\ref{eq:kjkk}) is expressed as
\begin{equation}
 \delta \sigma_{ k }=\sqrt {-\tau}\left[ { c }_{ 1 }{ H }_{ \nu  }^{ \left( 1 \right)  }\left( -k\tau  \right) +{ c }_{ 2 }{ H }_{ \nu  }^{ \left( 2 \right)  }\left( -k\tau  \right)  \right] \label{eq:kllll} ,
\end{equation}
where ${ H }_{ \nu\in C }^{ \left( 1 \right) }\left( x \right)$ and
${ H }_{ \nu\in C }^{ \left( 2 \right) }\left( x \right)$ are Hankel
functions of the first and second kind.\footnote{The Hankel functions of the first kind asymptotically behave as 
\begin{eqnarray} 
{ H }_{ \nu\in C  }^{ \left( 1 \right)  }\left( x\gg 1 \right) &\sim& \sqrt { \frac { 2 }{ \pi x }  } { e }^{ i\left( x-\frac {  \pi  }{ 2 } \nu-\frac { \pi  }{ 4 }  \right)  }, \nonumber \\
{ H }_{ \nu\in R }^{ \left( 1 \right)  }\left( x\ll 1 \right) &\sim&\left( -\frac{i}{\pi} \right)\Gamma \left( \nu \right)   
\left( \frac{1}{2}x\right)^{-\nu}, \nonumber \\
{ H }_{ \nu\in C  }^{ \left( 1 \right)  }\left( x\ll 1 \right) &\sim&\frac { i }{ \pi \nu  } \left[ { e }^{ -i\pi \nu  }\Gamma \left( 1-\nu \right) { \left( \frac { 1 }{ 2 } x \right)  }^{ \nu  }-\Gamma \left( 1+\nu \right) { \left( \frac { 1 }{ 2 } x \right)  }^{ -\nu  } \right]. 
\nonumber 
\end{eqnarray}
}  In order to determine the coefficients $c_{1}$ and
$c_{2}$, in the ultraviolet regime $ \left(-k\tau \gg 1\right)$ we match the solution with the postivie-frequency plane-wave solution in flat spacetime, $e^{-ik\tau }/\sqrt { 2k}$, which gives
\begin{equation}
c_{1}=\frac{\pi}{2}e^{i\left(\nu+\frac{1}{2}\right)\frac{\pi}{2}},\quad c_{2}=0. 
\end{equation}
The choice of a particular set of coefficients $c_{1}, c_{2}$ is equivalent to choosing the vacuum~\cite{Birrell:1982ix}.
On super-horizon scales $ \left(-k\tau \ll 1\right)$, the re-scaled mode functions of the Higgs take the form 
\begin{eqnarray}
 \delta \sigma_{ k }&=&\frac{\pi}{2}e^{i\left(\nu+\frac{1}{2}\right)\frac{\pi}{2}}\sqrt {-\tau}{ H }_{ \nu  }^{ \left( 1 \right)  }\left( -k\tau \right), \\ &=&e^{i\left(\nu-\frac{1}{2}\right)\frac{\pi}{2}}2^{\nu-\frac{3}{2}}\frac { \Gamma \left( \nu  \right)  }{ \Gamma \left( 3/2 \right)  } \frac { 1 }{ \sqrt { 2k }  } { \left( -k\tau  \right)  }^{ \frac { 1 }{ 2 } -\nu  }.
\end{eqnarray}

If we consider the case where the Higgs mass is light, i.e. $m_{\rm eff}\le 3H/2$, the absolute value of $\delta h_{ k }$ is
given as
\begin{eqnarray}
\left|\delta h_{ k }  \right| &=&\frac{H}{\sqrt{2k^{3}}}2^{\nu-\frac{3}{2}}\frac { \Gamma \left( \nu  \right)  }{ \Gamma \left( 3/2 \right)  } \frac { k }{ aH } { \left( \frac { k }{ aH }  \right)  }^{ \frac { 1 }{ 2 } -\nu  } , \nonumber \\
&\simeq&\frac{H}{\sqrt{2k^{3}}}{ \left( \frac { k }{ aH }  \right)  }^{ \frac { 3 }{ 2 } -\nu  }.
\end{eqnarray}
Integrating over super-horizon modes we obtain the variance of the Higgs field fluctuations a
\begin{eqnarray}  
 \left< { h }^{ 2 } \right> &=& \int _{ H  }^{ aH }{ {\left|\delta h_{ k }  \right|}^{2} \frac{d^{3}k}{\left(2\pi\right)^{3}}}, \\
 &\simeq&
     \frac { 3{ H }^{ 4 } }{ 8{ \pi  }^{ 2 }m_{\rm eff}^{ 2 } } \quad( m_{\rm eff}\ll 3H/2).\label{eq:hdffm}
\end{eqnarray}
Next we assume that the Higgs probability distribution function is Gaussian, i.e.
\begin{equation}
P\left( h,t \right) = \frac { 1 }{ \sqrt {2{ \pi  }\left< { h }^{ 2 } \right> } } \exp \left( -\frac { h^{2} }{ 2\left< { h }^{ 2 }\right> }  \right).\label{eq:hjfdg}
\end{equation}
By using Eq.~(\ref{eq:hsdsg}), the probability that the standard electroweak vacuum survives can be obtained as
\begin{eqnarray}
{ P }\left(  h<{ h }_{ \rm max },N_{\rm tot} \right) &\equiv& \int _{ -{ h }_{\rm  max } }^{ { h }_{ \rm max } }{ dh\ P\left( h,t_{\rm end}\right)  }, \\ &=& {\rm erf}\left( \frac{{ h }_{ \rm max }  }{ \sqrt { 2\left< { h }^{ 2 } \right> }  }  \right).
\end{eqnarray}
On the other hand, the probability that the Higgs falls into the true vacuum is expressed as
\begin{eqnarray}
{ P }\left(  h>{ h }_{ \rm max },N_{\rm tot} \right)&=&1-{\rm erf}\left( \frac{{ h }_{ \rm max }  }{ \sqrt { 2\left< { h }^{ 2 } \right> }  }  \right), \\ &\simeq& \sqrt {\frac {2}{\pi}}\frac {\sqrt{\left< { h }^{ 2 } \right> }}{{ h }_{ \rm max }}e^{-\frac {{ h }_{ \rm max }^{2}}{2\left< { h }^{ 2 } \right>}}.
\end{eqnarray}
Imposing the condition shown in (\ref{eq:horsdsd}),
we obtain the relation
\begin{equation}
\frac{\left< { h }^{ 2 } \right>}{{ h }_{ \rm max }^{2}}<\frac{1}{6N_{\rm hor} }\label{eq:hssssdg}.
\end{equation}
If we substitute $\langle h^2\rangle$ from Eq.~(\ref{eq:hdffm}) into this relation, we find the upper bound on $H$ to be
\begin{equation}
\frac { H }{ { h }_{ \rm max } } < \frac { 2{ \pi  }m_{\rm eff} }{ 3H\sqrt{N_{\rm hor} }}\label{eq:nndfdf},
\end{equation}
which is the same as the constraint given in
Ref.\cite{Espinosa:2015qea}. We plot this line in Fig.~\ref{Fig:2},
where we assume $h_{\rm max}\sim \Lambda_{I}$ because of the small nonminimal
coupling $\xi$ and it is labelled by ``Inflation Stage''.

%%%%%%%%%%%%%%%%%%%%%%%%%%%%%%%%%%%%%%%%%%%%%%%%%%%%%%%%%%%%%%%%%%%%%%
\subsection{Fluctuations of Massive Higgs field}
%%%%%%%%%%%%%%%%%%%%%%%%%%%%%%%%%%%%%%%%%%%%%%%%%%%%%%%%%%%%%%%%%%%%%%
In this subsection, we consider the case of a large effective Higgs mass, namely
$m_{\rm eff} >3H/2$.  We define $\tilde {\nu }$  as
\begin{equation}
\tilde {\nu  }=\sqrt { \frac{m_{\rm eff}^{2}}{H^{2}} -\frac{9}{4}}\label{eq:ksd}.
\end{equation}
On super-horizon scales, the re-scaled Higgs fluctuations are given by
\begin{equation}
\delta \sigma_{ k }=\frac{\pi}{2}e^{i\left(i\tilde { \nu  } +\frac{1}{2}\right)\frac{\pi}{2}}\sqrt {-\tau}{ H }_{ i\tilde { \nu  }  }^{ \left( 1 \right)  }\left( -k\tau \right).
\end{equation}
The absolute value of $\delta \sigma_{ k }$ is obtained as
\begin{equation}
 \left|\delta \sigma_{ k }  \right| \simeq \frac { 1  }{ \sqrt {2k} } \left( -k\tau\right)^{1/2}\frac { { e }^{ \frac { \pi  }{ 2 } \tilde { \nu  }  } }{ \sqrt {2\pi}\tilde { \nu  }} \left| \Gamma \left( 1-i\tilde { \nu  }  \right) \right|.
\end{equation}
Using the relation
${ \left| \Gamma \left( 1-iy \right) \right| }^{ 2 }=\pi y/\sinh {
  \left( \pi y \right) }$,
the fluctuations of the massive Higgs field are estimated to be
\begin{eqnarray}
\left|\delta h_{ k }  \right|^{2} \simeq \frac{1}{2a^{3}\tilde { \nu  } H}.
\end{eqnarray}
As such, the variance of the vacuum fluctuations during inflation is given as 
\begin{eqnarray}
\left< { h }^{ 2 } \right> &=& \int _{ H  }^{ aH }{ {\left|\delta h_{ k }  \right|}^{2} \frac{d^{3}k}{\left(2\pi\right)^{3}}}
\simeq \frac { H^{2}}{ 12{ \pi  }^{ 2 }\tilde { \nu  }  },\\
&\approx&  \frac { H^{3}}{ 12{ \pi  }^{ 2 }m_{\rm eff} }.\label{eq:kksds}
\end{eqnarray}
Inflationary effective mass terms thus lift the effective Higgs potential
 and suppress the Higgs vacuum fluctuations.  Substituting the above result into \eqref{eq:hssssdg}, in the case of a massive Higgs field the requirement that our observable Universe contains no AdS domains gives us the condition
\begin{equation}
\frac { H }{ { h }_{ \rm max } } <\sqrt { \frac { 2{ \pi  }^{ 2 }{ m }_{ \rm eff } }{ H{ N }_{\rm  hor } }  } \label{eq:haasdg}.
\end{equation}

The constraint on the nonminimal coupling $\xi$ can be estimated by 
those conditions shown in (\ref{eq:nndfdf}) and (\ref{eq:haasdg}).
 If we assume
$m_{\rm eff}=\sqrt{12\xi}H$, we obtain a lower bound on the
nonminimal coupling as $\xi>0.03$ using the fact that $h_{\rm max}\simeq 10 m_{\rm eff}$.\footnote{The total Higgs potential during inflation can be approximated by
\[ V_{\rm eff }\left( h \right) \simeq \frac { 1 }{ 2 } { m }_{ \rm eff }^{ 2 }{ h }^{ 2 }\left( 1-\frac { 1 }{ 2 } { \left( \frac { h }{ { h }_{\rm  max } }  \right)  }^{ 2 } \right), \] 
where ${ h }_{\rm  max }$ is expressed to be
\[{ h }_{ \rm max }=\sqrt { -\frac { { m }_{ \rm eff }^{ 2 } }{ {
        \lambda }_{ \rm eff } } }. \]
Our assumption $h_{\rm max}\simeq 10 m_{\rm eff}$ is numerically valid
for the RG-improved effective potential.}
This constraint corresponds to the vertical line in Fig.~\ref{Fig:2}.
In the case of inflaton-Higgs coupling, where $m_{\rm eff} = g\phi$, 
we can similarly obtain a constraint on $g$, but it will depend
on the inflaton field value $\phi$ and the Hubble scale $H$.
 
Whilst inflationary effective masses can prevent the Higgs from
evolving into the true vacuum during inflation, after inflation they
become ineffective, and the Higgs field fluctuations generated as a
result of resonant preheating may destabilize the standard electroweak
vacuum. We will discuss this problem in the next section.

%%%%%%%%%%%%%%%%%%%%%%%%%%%%%%%%%%%%%%%%%%%%%%%%%%%%%%%%%%%%%%%%%%%%%%
\section{Higgs fluctuations after inflation and during the preheating stage}
%%%%%%%%%%%%%%%%%%%%%%%%%%%%%%%%%%%%%%%%%%%%%%%%%%%%%%%%%%%%%%%%%%%%%%
After the end of inflation, the inflaton field $\phi$ oscillates near
the minimum of its potential and produces a huge amount of
elementary particles that interact with each other and eventually form
a thermal plasma.  The reheating process is generally classified into
several stages.  In the first stage, the classical,
coherently-oscillating inflaton field $\phi$ may give rise to the
production of massive bosons due to parametric resonance.  In most
cases, this first stage occurs extremely rapidly. This
nonthermal period is called preheating~\cite{Kofman:1997yn}, and is
different from the subsequent stages of reheating and
thermalization.  Parametric resonance in the preheating stage
may sometimes produce topological defects or lead to nonthermal phase
transitions~\cite{Kofman:1995fi}.

In Ref.\cite{Herranen:2015ima}, the authors discussed the resonant
production of Higgs fluctuations after inflation in the case that the
Higgs is non-minimally coupled to gravity.  However, the preheating
dynamics is extremely complicated, and it is difficult to estimate
analytically the Higgs vacuum fluctuations during the preheating
stage.\footnote{The authors of Ref.\cite{Ema:2016kpf} gave a comprehensive study of 
the parametric resonance of the nonminimal coupling $\xi$ 
or the inflaton-Higgs coupling $g$ by using the lattice simulations, 
and their results are consistent with ours.}
In this section, we numerically analyse the Higgs fluctuations
after inflation and during the preheating stage.  After inflation, the
re-scaled Higgs mode solution is no longer given by
Eq.~(\ref{eq:kllll}).\footnote{In a de Sitter background, the
Klein-Gordon equation for $\delta\sigma_k$ takes the form
  \[ \delta \sigma_{ k
    }''+\left(k^{2}-\frac{1}{\tau^{2}}\left(2-\frac{m_{\rm
            eff}^{2}}{H^{2}}\right)\right) \delta \sigma_{ k }=0. \]
  However, during the preheating period, if we assume that the inflaton potential is quadratic then the Universe behaves like that of a matter-dominated Universe, in which case the Klein-Gordon equation takes the form
  \[\delta \sigma_{ k }''+\left(k^{2}+m_{\rm
        eff}^{2}\tau^{4}-\frac{2}{\tau^{2}}\right) \delta \sigma_{ k
    }=0. \]}
Instead, we use the WKB approximation and obtain the variance of 
the massive Higgs fluctuations which correspond with 
the result given by Eq.~(\ref{eq:kksds}).  

The Klein-Gordon equation for $k$-modes of the Higgs field is
given as
\begin{equation}
{ \delta \ddot { h }  }_{ k }+3H{ \delta \dot { h }  }_{ k }+\left( \frac{k^{2} }{a^{2}}+m_{\rm eff}^{2}\right){ \delta { h }  }_{ k }=0,
\end{equation}
which can be re-written in the useful form
\begin{equation}
\begin{split}
\frac {d^{2}\left(a^{3/2} {\delta { h }  }_{ k } \right)}{  dt^{2}}+\left(\frac{k^{2} }{a^{2}}+m_{\rm eff}^{2}-\frac{9}{4}H^{2}-\frac{3}{2}\dot {  H}  \right)
\left(a^{3/2} {\delta { h }  }_{ k } \right)=0.
\end{split}
\end{equation}
If we consider the massive Higgs field case, i.e. $m_{\rm eff}>3H/2$,
then the Higgs mode functions are given by
\begin{equation}
{ \delta  { h }  }_{ k }\simeq \frac{e^{-i\omega_{k}\left(t\right) \cdot t}}{a^{3/2}\sqrt{2\omega_{k}\left(t\right)}}.
\end{equation}
where $\omega_{k}^{2}\simeq\frac{k^{2} }{a^{2}}+m_{\rm eff}^{2}$ and we
have assumed the adiabatic condition $\dot { \omega}_{k}\left(t\right)/
\omega_{k}^{2}\left(t\right)\ll 1$ is satisfied, and ${
\omega}_{k}^{2}\left(t\right)>0$.  As such, the amplitude $\left| {
\delta { h } }_{ k } \right|^{2} $ after inflation is estimated to
be~\cite{Jin:2004bf}
\begin{equation}
\left| { \delta  { h }  }_{ k } \right|^{2} \simeq \frac{1}{2a^{3}\omega_{k}}.
\end{equation}
Hence, the variance of the massive Higgs fluctuations which are outside
the Hubble radius after inflation is given as
\begin{eqnarray}
\left< { h }^{ 2 } \right>_{\rm end} &=& \frac{1}{2\pi^{2}}\int _{ 0 }^{ a_{\rm end}H_{\rm end}  }k^{2}{ {\left|\delta h_{ k }  \right|}^{2}dk}, \\
&\simeq& \frac { H_{\rm end}^{3}}{ 12{ \pi  }^{ 2 }m_{\rm eff} }.\label{eq:kbbb}
\end{eqnarray}
This can be used as an estimate for the minimum amplitude of the
homogeneous Higgs field after inflation. The above Higgs fluctuations are consistent with
the result given by Eq.~(\ref{eq:kksds})
\footnote{Note that if we consider the light Higgs field case, i.e. $m_{\rm eff}<3H/2$,
the Higgs vacuum fluctuations at the end of inflation are consistent with
the result given by Eq.~(\ref{eq:hdffm}).} and exponentially amplified by parametric resonance.  
If we substitute Eq.~(\ref{eq:kbbb}) into (\ref{eq:hssssdg}), we obtain the constraint
\begin{equation}
\frac { H_{\rm end} }{ \Lambda_{I} } < \sqrt { \frac { 2\pi^{2}  m_{\rm eff}}{ N_{\rm hor}H_{\rm end}} }\label{eq:kbdfdfbb}.
\end{equation}
Note that the effective mass ( $m_{\rm eff}\simeq\sqrt {\xi R\left(t\right)}$~\footnote{ The scalar curvature $R\left(t\right)$ is written as
  \[R\left( t \right) =\frac { 1 }{ { M }_{\rm  pl }^{ 2 } } \left[
      4V\left( \phi \right) -{ \dot { \phi } }^{ 2 } \right] .\]}
or $m_{\rm eff}\simeq g\phi\left(t\right)$) decreases and sometimes
disappears during the preheating period. Therefore, 
the effective mass cannot stabilizes the effective Higgs potential
$V_{\rm eff}\left(h\right)$, and we can assume $h_{\rm max}\simeq \Lambda_{I}$.

Let us consider the amplification of the Higgs vacuum fluctuations via
parametric resonance.  For simplicity, we consider chaotic inflation
with a quadratic potential as an example, i.e.
\begin{equation}
V_{\rm inf}\left( \phi \right)=\frac{1}{2}m_{\phi}^{2}\phi^{2},
\end{equation}
where $m_{\phi}\simeq 7\times10^{-6}M_{\rm pl}$. In the chaotic inflation
scenario, inflation occurs at super-Planckian field values, $\phi>5M_{\rm
pl}$. Primordial density perturbations relevant for the CMB are produced
at around $\phi\sim 15M_{\rm pl}$, and inflation terminates at $\phi\sim3
M_{\rm pl}$. After inflation, the inflaton field oscillates as
\begin{eqnarray}
\phi \left(t\right)&=& \Phi \left(t\right)\sin { m_{\phi}t},\\
\Phi\left(t\right)&=&\sqrt{\frac{8}{3}}\frac{M_{\rm pl}}{m_{\phi}t}.
\end{eqnarray} 
When the inflaton field oscillates, the effective masses of the fluctuations of 
$h$ evolve in a highly non-adiabatic way, which leads to them being produced 
explosively via parametric resonance.

The Klein-Gordon equation for the Higgs field given in Eq.~(\ref{eq:kdfd})
can be rewritten as
\begin{equation}
\begin{split}
\frac {d^{2}\left(a^{3/2} {\delta { h }  }_{ k } \right)}{  dt^{2}}+\left(\frac{k^{2} }{a^{2}}+g^{2}\phi^{2}+
\frac{1}{M_{\rm pl}^{2}} \left(\frac{3}{8}-\xi \right)\dot { \phi }\right. \\ \left.-\frac{1}{M_{\rm pl}^{2}} \left(\frac{3}{4}-4\xi \right)V\left( \phi \right) \right)
\left(a^{3/2} {\delta { h }  }_{ k } \right)=0\label{eq:hfhfff}.
\end{split}
\end{equation}
Eq.~(\ref{eq:hfhfff}) can be reduced to the well-known Mathieu equation as
follows
\begin{equation}
\frac {d^{2}\left(a^{3/2} {\delta { h }  }_{ k } \right)}{  dz^{2}}+\left(A_{k}-2q\cos2z\right)
\left(a^{3/2} {\delta { h }  }_{ k } \right)=0,\label{mathieu}
\end{equation}
where $z=m_{\phi}t$ and $A_{k}$ and $q$ are given as 
\begin{eqnarray}
A_{k}&=&\frac{k^{2}}{a^{2}m_{\phi}^{2}}+\frac{g^{2}\Phi^{2}\left(z\right)}{2m_{\phi}^{2}}+\frac{\Phi^{2}\left(z\right)}{2 M_{\rm pl}^{2}}\xi, \\
q&=&\frac{3\Phi^{2}\left(z\right)}{4M_{\rm pl}^{2}}\left(\xi-\frac{1}{4}\right)+\frac{g^{2}\Phi^{2}\left(z\right)}{4m_{\phi}^{2}}.
\end{eqnarray}
The properties of the solutions to the Mathieu equation can be
classified using a stability/instability chart. The solutions of the
Mathieu equation show broad resonance when $q\gg 1$ or narrow resonance
when $q<1$. In the context of preheating, $A_{k}$ and $q$ are dependent
on $z$ due to the expansion of the Universe, making it very difficult to
derive analytical solutions. However, we can roughly estimate $ {\delta
{ h } }_{ k }$ by using the Floquet exponent $\mu_{k}$. In the broad
resonance regime, where $q\gg 1$, parametric resonance amplifies the
Higgs vacuum fluctuation after inflation, giving~\cite{Liddle:1999hq,Kofman:1997yn}
\begin{equation}
\left< { h }^{ 2 } \right>=\left< { h }^{ 2 } \right>_{\rm end}\ e^{2\pi\mu_{k}m_{\phi} t} \left( \frac { m_{\rm eff}\left(t_{\rm end}\right) }{ m_{\rm eff}\left(t \right) }  \right){ \left( \frac { H\left(t\right) }{ { H }_{ \rm end } }  \right)  }^{ 3 } ,\label{growth}
\end{equation}
where the Floquet exponent $\mu_{k}$ is given as
\begin{equation}
\mu_{k}\simeq \frac { 1 }{ 2\pi  } \ln { \left( 1+2{ e }^{ -\pi { \kappa  }^{ 2 } } \right)}, \quad
{ \kappa  }^{ 2 }=\frac { { A }_{ k }-2q }{ 2\sqrt { q }  }.
\end{equation}
We can take
${ \kappa }^{ 2 } \ll 1$ for all modes outside the horizon 
scale after inflation. Then we obtain
$\mu_{k}\simeq \frac { 1 }{ 2\pi } \ln { 3} \simeq 0.17$.
The broad resonance requires $q\gg 1$. Therefore, the period
of the broad resonance is
$m_{\phi}t\ll \sqrt {
  \frac{3}{4}\left(\xi-\frac{1}{4}\right)+\frac{g^{2}M_{\rm
      pl}^{2}}{4m_{\phi}^{2}}}$.
Then narrow resonance follows the broad resonance.

%%%%%%%%%%%%%%%%%%%%%%%%%%%%%%%%%%%%%%%%%%%%%%%%%%%%%%%%%%%%%%%%%%%%%%
\begin{figure}[t]
\includegraphics[width=87mm]{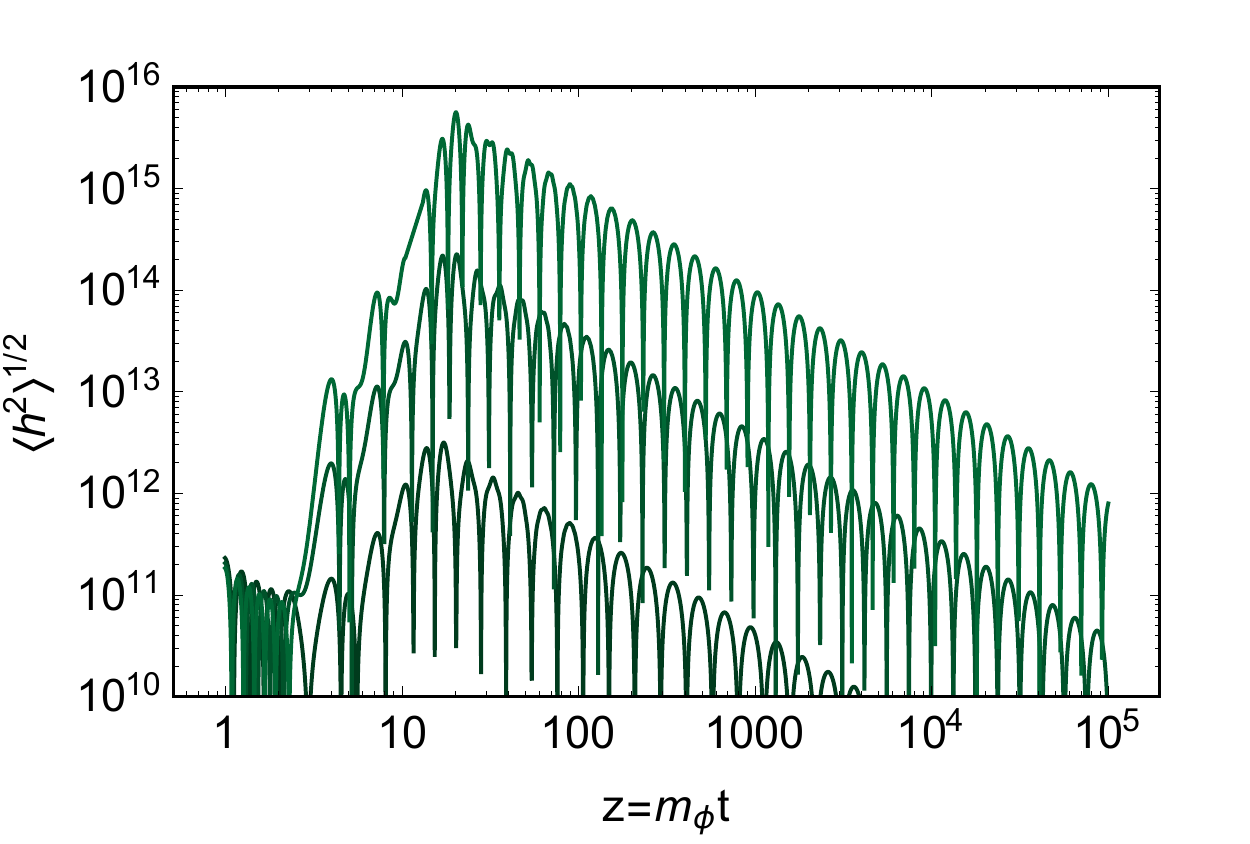} \caption{Higgs vacuum fluctuation
variance in the $\xi$-resonance scenario. In the lower, middle and upper
curves we have used the nonminimal couplings $\xi=10^{1.4}$, $\xi
=10^{1.6}$ and $\xi = 10^{1.8}$ respectively. Broad resonance occurs
strongly for $\xi>10^{1.6}$.}  \label{Fig:3}
\end{figure}
%%%%%%%%%%%%%%%%%%%%%%%%%%%%%%%%%%%%%%%%%%%%%%%%%%%%%%%%%%%%%%%%%%%%%%

%%%%%%%%%%%%%%%%%%%%%%%%%%%%%%%%%%%%%%%%%%%%%%%%%%%%%%%%%%%%%%%%%%%%%%
\begin{figure}[t]
\includegraphics[width=87mm]{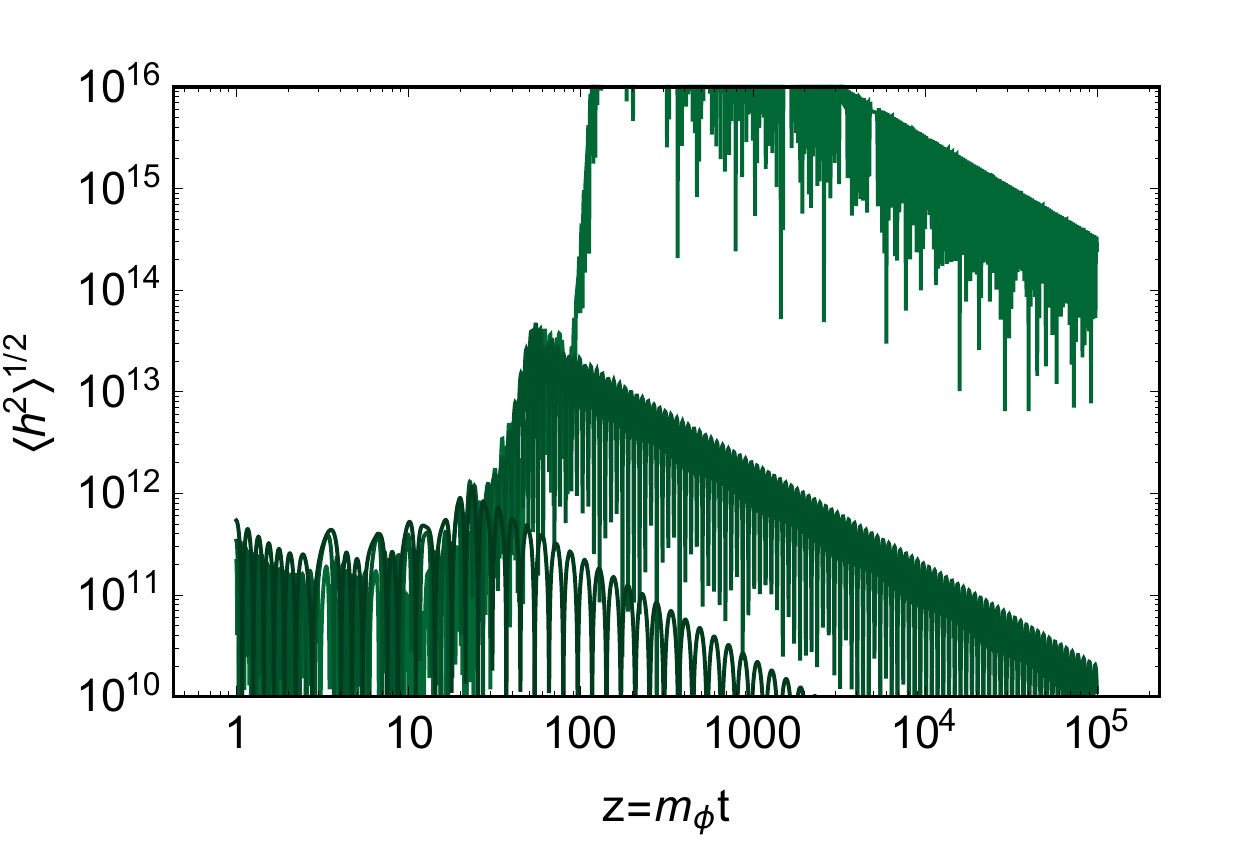} \caption{Higgs vacuum fluctuation
variance in the $g$-resonance scenario, where we ignore back
reaction effects. In the lower, middle and upper curves we have used the
inflaton-Higgs coupling $g=10^{-4.4},10^{-4}$ and $10^{-3.6}$
respectively. Broad resonance occurs strongly for
$g>10^{-4}$.}\label{Fig:4}
\end{figure}
%%%%%%%%%%%%%%%%%%%%%%%%%%%%%%%%%%%%%%%%%%%%%%%%%%%%%%%%%%%%%%%%%%%%%%

%%%%%%%%%%%%%%%%%%%%%%%%%%%%%%%%%%%%%%%%%%%%%%%%%%%%%%%%%%%%%%%%%%%%%%
\subsection{Parametric resonance via nonminimal gravity-Higgs coupling $\xi$}
%%%%%%%%%%%%%%%%%%%%%%%%%%%%%%%%%%%%%%%%%%%%%%%%%%%%%%%%%%%%%%%%%%%%%%
In this subsection, we solve numerically the Mathieu equation in the
  case of geometric preheating~\cite{Bassett:1997az,Tsujikawa:1999jh},
  where $m_{\rm eff}^2$ is dominated by the $\xi R$ term. We call this
  the $\xi$-resonance scenario.\footnote{In the $\xi$-resonance
  scenario, in order to obtain parametric resonance we require $\xi\gg
  1$ or $\xi<0$. The parametric amplification obtained for $\xi<0$ is
  extremely strong compared with that obtained for $\xi\gg
  1$~\cite{Bassett:1997az,Tsujikawa:1999jh}, but here we only consider positive $\xi$, as we are interested in the case where the effective mass acts so as to stabilize the Higgs during inflation.}  In this case $A_{k}$ and $q$ are
  given by
\begin{equation}
A_{k}\simeq\frac{k^{2}}{a^{2}m^{2}}+\frac{\Phi^{2}\left(z\right)}{2M_{\rm pl}^{2}}\xi ,\quad
q\simeq\frac{3\Phi^{2}\left(z\right)}{4M_{\rm pl}^{2}}\left(\xi-\frac{1}{4}\right).
\end{equation}
In Fig.~\ref{Fig:3} we show the evolution of the variance of the Higgs
vacuum fluctuations as obtained by numerically solving
eq.~\eqref{mathieu} with $A_k$ and $q$ as given above.  We take three
different values for the nonminimal coupling, namely
$\xi=10^{1.4},10^{1.6}$ and $10^{1.8}$. We find that broad resonance can
occur strongly for $\xi>10^{1.6}$.  We see that the nonminimal coupling
$\xi$ is constrained to be $\xi<10^{1.6}$ in order not to produce the
Higgs AdS domains via the parametric resonance.

%%%%%%%%%%%%%%%%%%%%%%%%%%%%%%%%%%%%%%%%%%%%%%%%%%%%%%%%%%%%%%%%%%%%%%
\subsection{Parametric resonance via inflaton-Higgs coupling $g$}
%%%%%%%%%%%%%%%%%%%%%%%%%%%%%%%%%%%%%%%%%%%%%%%%%%%%%%%%%%%%%%%%%%%%%%
In this subsection we solve numerically the Mathieu equation in the
standard preheating scenario, where $m_{\rm eff}^2$ is dominated by the
$g^2\phi^2$ term.  We call this the $g$-resonance scenario. In this case
$A_{k}$ and $q$ are given by
\begin{equation}
A_{k}\simeq\frac{k^{2}}{a^{2}m_{\phi}^{2}}+\frac{g^{2}\Phi^{2}\left(z\right)}{2m_{\phi}^{2}},\quad
q\simeq\frac{g^{2}\Phi^{2}\left(z\right)}{4m_{\phi}^{2}}.
\end{equation}
In Fig.~\ref{Fig:4} we show the evolution of the variance of the Higgs
vacuum fluctuations as obtained by numerically solving \eqref{mathieu}
with $A_k$ and $q$ given as above.  For the sake of simplicity, here we
have neglected back reaction effects, which we comment further on
below. We consider three different values for the inflaton-Higgs
coupling, namely $g=10^{-4.4},10^{-4}$ and $10^{-3.6}$. Broad resonance
can occur strongly for $g>10^{-4}$. However, $g$ is restricted in order
not to give rise to large radiative corrections to the inflaton
potential~\cite{Linde:2007fr},
\begin{equation}
\Delta V_{\rm inf}\simeq \frac { { g }^{ 4 } }{ 64{ \pi  }^{ 2 } } { \phi  }^{ 4 }\log { \frac { g^{2} { \phi  }^{ 2 } }{ m_{ \phi  }^{2}  } }.
\end{equation}
Thus, for quadratic chaotic inflation-type models, the inflaton-Higgs
coupling is constrained to be $g<10^{-3}$.
Because $g$-resonance can occur in the parameter range $10^{-4}<g<10^{-3}$, we
find an upper bound on $g$ as $g<10^{-4}$.

In the early stages of parametric resonance, our semiclassical approximation is
valid. On the other hand, in the later stages, backreaction effects and
effects of scattering among the created particles become
important (see, e.g. Ref.\cite{Boyanovsky:1996sq}). 
In this case, our semiclassical approximation may break down.
However, well before the backreaction effects become significant, the
generated fluctuations of the Higgs field immediately grow and exceeds
the hill of the effective potential due to the parametric resonance.
Actually, we can estimate the maximal Higgs fluctuation 
$\left< { h }^{ 2 } \right>\sim\left< { \phi }^{ 2 } \right>\sim m_{\phi}^{2}/g^{2}$
where backreaction effects terminate the amplification of the Higgs fluctuations.
The maximal Higgs fluctuation can be estimated as
$\sqrt {\left< { h }^{ 2 } \right> } \sim m_{\phi}/g \sim 10^{17}\ {\rm GeV}$ where
$m_{\phi}\simeq 7\times10^{-6}M_{\rm pl}$ and $g\sim 10^{-4}$
and the generated Higgs fluctuations immediately 
overcomes the instability scale $\Lambda_{I}$ 
before the backreaction effects become significant.
Therefore, the backreaction effects cannot make a significant contribution
to the electroweak vacuum stability.

In Fig.~\ref{Fig:2}, we plot the upper bounds
on $H/h_{\rm max}$ as a function of the nonminimal gravity-Higgs
coupling $\xi$. We have assumed that the effective mass is $m_{\rm
eff}=\sqrt{12\xi}H$.  In the left panel, we do not include the
constraint on $\xi$ coming from parametric amplification of the Higgs during preheating, as this constraint is model dependent, i.e. it depends on how the inflaton behaves after inflation. On the other hand, in the right
panel we have included the constraint on $\xi$ arising from broad resonance during the
preheating stages.  For simplicity, here we assume the quadratic chaotic
inflation model with $V_{\rm inf}\left( \phi
\right)=\frac12m_{\phi}^{2}\phi^{2}$, which gives us the constraint $\xi<10^{1.6}$.

%%%%%%%%%%%%%%%%%%%%%%%%%%%%%%%%%%%%%%%%%%%%%%%%%%%%%%%%%%%%%%%%%%%%%%
\section{Thermal fluctuations during the reheating era}
%%%%%%%%%%%%%%%%%%%%%%%%%%%%%%%%%%%%%%%%%%%%%%%%%%%%%%%%%%%%%%%%%%%%%%
During the reheating stage, most of the inflaton energy is transferred
to the thermal energy of elementary particles. The reheating process
finishes approximately when $H = \Gamma_{\rm tot}$. Therefore, the reheating
temperature can be expressed as
\begin{equation}
T_{\rm reh}=\left(\frac{90}{\pi^{3}g_{*}}\right)^{1/4}\sqrt { M_{\rm pl} \Gamma_{\rm tot}},
\end{equation} 
where $g_{*}$ is the number of relativistic degrees of freedom.
\footnote{
When the effective mass of the inflaton field $\phi$ is too small to start to
oscillate,  the thermal bath is not produced. Then there would be a
time lag for the  production of the thermal bath until the beginning
for the oscillation of the inflaton field $\phi$~\cite{Kamada:2014ufa}.  
In this case, we can adopt the constraints obtained in \eqref{eq:kbdfdfbb}.}
%%%%%%%%%%%%%%%%%%%%%%%%%%%%%%%%%%%%%%%%%%%%%%%%%%%%%%%%%%%%%%%%%%%%%%%%
\begin{figure}[t]
\includegraphics[width=87mm]{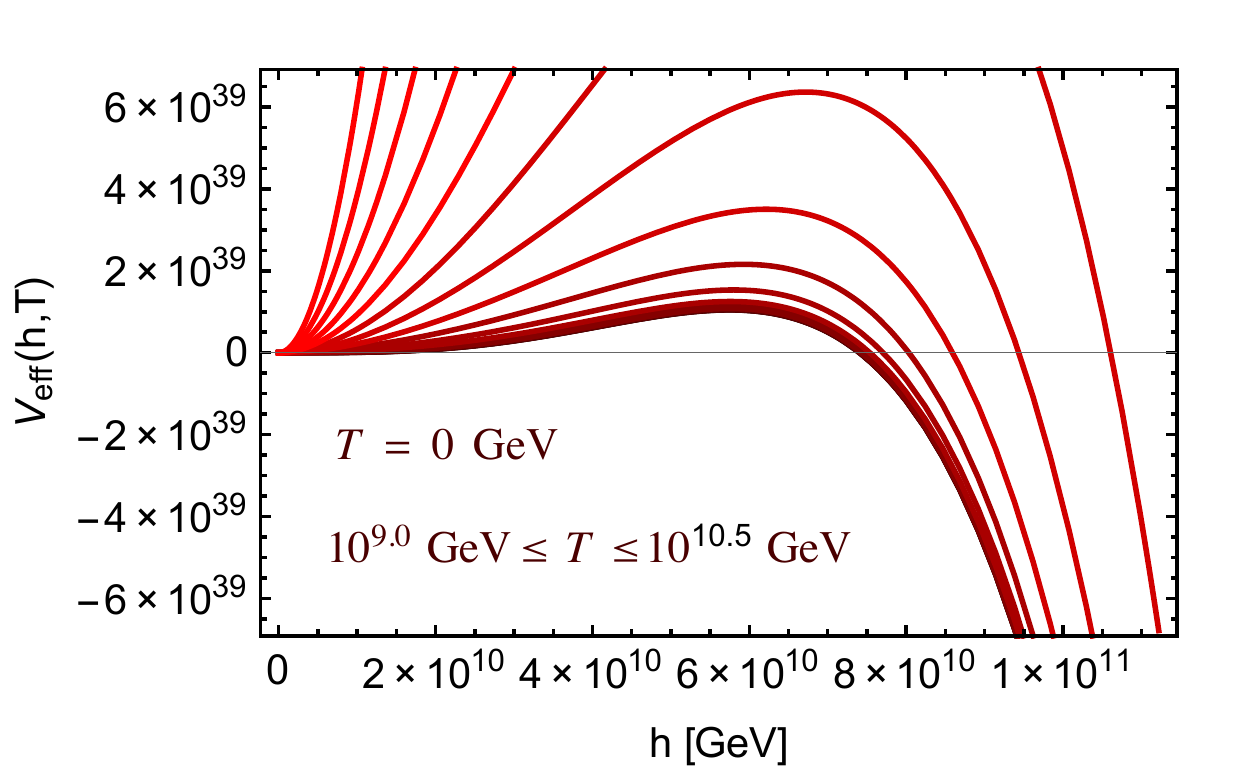}
\caption{RG improved effective Higgs potential at finite temperature for $T=0\ {\rm GeV}$ and $10^{9.0}\ {\rm GeV}\le T \le 10^{10.5}\ {\rm GeV}$ on the present best-fit values of $m_{h}$ and $m_{t}$. }
\label{Fig:5}
\end{figure}
%%%%%%%%%%%%%%%%%%%%%%%%%%%%%%%%%%%%%%%%%%%%%%%%%%%%%%%%%%%%%%%%%%%%%%%%

Thermal effects in the reheating can raise the effective potential of the Higgs field.
The RG improved effective Higgs potential at finite temperature
is given by the familiar zero-temperature corrections and the thermal corrections as
\begin{equation}
V_{\rm eff}\left( h,T \right)=V_{\rm eff}\left( h\right)+\Delta { V }_{\rm eff}\left( h,T \right),
\end{equation}
The one-loop thermal corrections to the effective Higgs potential is
given as~\cite{Carrington:1991hz,Anderson:1991zb,Delaunay:2007wb},
\begin{align}
\Delta { V }_{\rm eff}\left( h,T \right)&=\sum _{ i=W,Z,t }{ \frac { { n }_{ i }{ T }^{ 4 } }{ 2{ \pi  }^{ 2 } } \int _{ 0 }^{ \infty  }{ dk{ k }^{ 2 }\ln { \left( 1\mp { e }^{ -\sqrt { { k }^{ 2 }+\frac{{ m }_{ i }^{ 2 }\left( h \right)}{ { T }^{ 2 } }}  } \right)  }    }  }  \nonumber \\
&=\sum _{ i=W,Z}{ { n }_{ i } J_{B}\left( m_{i}, T \right)}+
\sum _{ i=t}{  { n }_{ i } J_{F}\left( m_{i}, T \right)}.
\end{align}
Here we concentrate on the contributions from W bosons, Z bosons and top
quarks. $J_{B}$ ($J_{F}$) is the thermal bosonic
(fermionic) function, $m_{i}\left( h \right)$ is the background-dependent mass of $W$, $Z$ and $t$.

%%%%%%%%%%%%%%%%%%%%%%%%%%%%%%%%%%%%%%%%%%%%%%%%%%%%%%%%%%%%%%%%%%%%%%
\begin{figure}[t]
\includegraphics[width=87mm]{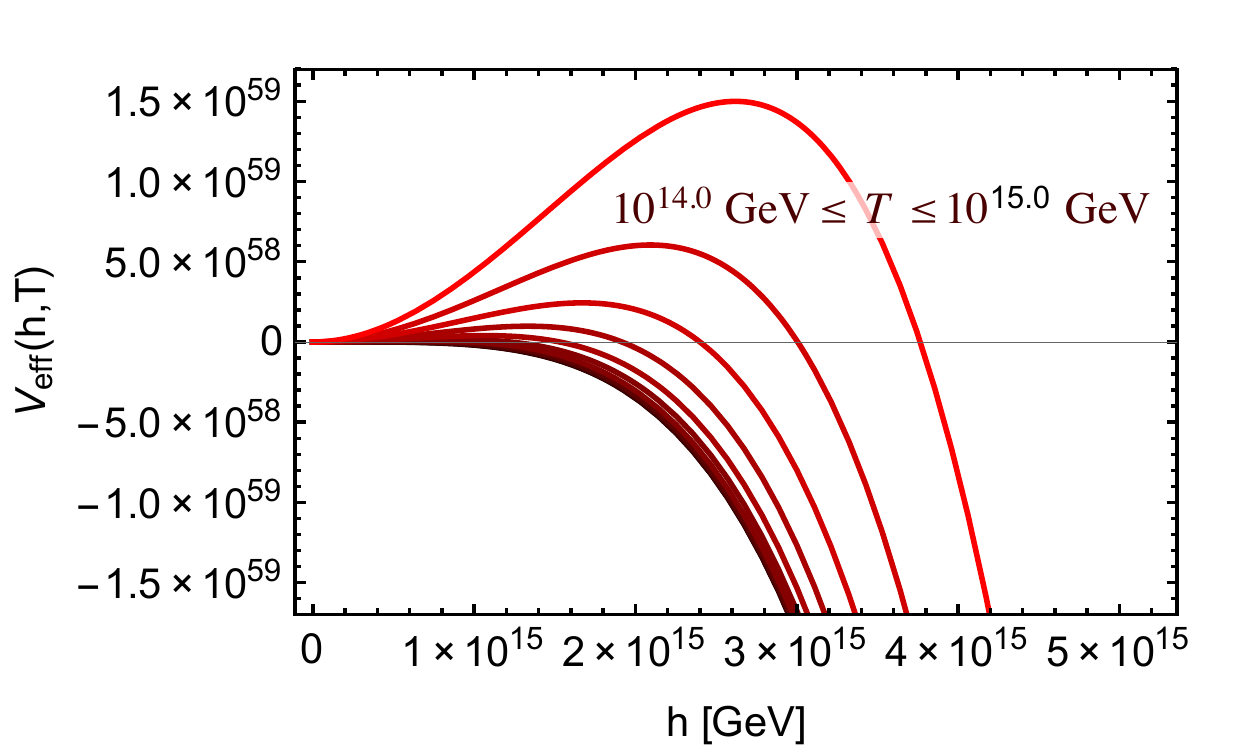}
\caption{RG improved effective Higgs potential at finite temperature
  for $10^{14.0}\ {\rm GeV}\le T \le 10^{15.0}\ {\rm GeV}$.
  The maximum of the Higgs potential is $h_{\rm max}=2.62\ T$ for $T=10^{15.0}\ {\rm GeV}$. }
\label{Fig:6}
\end{figure}
%%%%%%%%%%%%%%%%%%%%%%%%%%%%%%%%%%%%%%%%%%%%%%%%%%%%%%%%%%%%%%%%%%%%%%

In Fig.~\ref{Fig:5} and Fig.~\ref{Fig:6}, we plot the RG improved 
effective Higgs potential at finite temperature for a range of temperatures. In Fig.~\ref{Fig:5},
we plot the potential for $T=0\ {\rm GeV}$ and
$10^{9.0}\ {\rm GeV}\le T \le 10^{10.5}\ {\rm GeV}$. 
In Fig.~\ref{Fig:6}, we plot the potential for
$10^{14.0}\ {\rm GeV}\le T \le 10^{15.0}\ {\rm GeV}$. From the
figures, we see that although the high-temperature effects raise the
effective potential, it cannot be stabilized up to high energy
scales unless new physics emerges below the Planck scale.
Therefore, if the coherent Higgs field get over $h_{\rm max}$ during inflation or
preheating stage, the generated coherent Higgs field cannot go back to the 
electroweak vacuum by the high temperature effects.
 
In the high-temperature limit ($T\gg { m }_{ i }$), the thermal bosonic (fermionic) function $J_{B}$ ($J_{F}$) 
can be approximately written as
\begin{eqnarray}
J_{B}\left( m_{i}, T \right)&\simeq&-\frac{\pi^{2}T^{4}}{90}+\frac{m_{i}^{2}T^{2}}{24}-\frac{m_{i}^{3}T}{12\pi}-\frac { { m }_{ i }^{ 4 } }{ 64{ \pi  }^{ 2 } } \log { \frac { { m }_{ i }^{ 2 } }{ { a }_{ B }{ T }^{ 2 } }  },\nonumber \\ 
J_{F}\left( m_{i}, T \right)&\simeq&\frac{7}{8}\frac{\pi^{2}T^{4}}{90}-\frac{m_{i}^{2}T^{2}}{48}-\frac { { m }_{ i }^{ 4 } }{ 64{ \pi  }^{ 2 } } \log { \frac { { m }_{ i }^{ 2 } }{ { a }_{ F }{ T }^{ 2 } }  },
\end{eqnarray}
where $\log{a_{B}}\simeq5.408$ and $\log{a_{F}}\simeq2.635$. Here we
omit the terms which are independent of $h$. As such, the one-loop thermal 
corrections to the effective Higgs potential in the high-temperature limit  ($T\gg { m }_{ i }$) 
is approximately written as
\begin{eqnarray}
\Delta { V }_{\rm eff}\left( h,T \right) \simeq \frac{1}{2}c_{T}T^{2}h^{2}+\frac{1}{3}d_{T}T h^{3}+\frac{1}{4}\lambda_{T}h^{4},
\end{eqnarray}
where
\begin{eqnarray}
c_{T}&=&\frac{3g^{2}+g'^{2}+4y_{t}^{2}}{16},\ d_{T}=\frac{6g^{3}+3\left(g^{2}+g'^{2}\right)^{3/2}}{32\pi}, \\
\lambda_{T}&=&\scalebox{0.73}{$\displaystyle \frac { 3 }{ 64{ \pi  }^{ 2 } } \left( -\frac { { g }^{ 4 } }{ 2 } \log { \frac { { m }_{ W }^{ 2 }\left( h
\right)  }{ { a }_{ B }{ T }^{ 2 } }  } -\frac { { \left( { g }^{ 2 }+{ g' }^{ 2 } \right)  }^{ 2 } }{ 4 } \log { \frac { { m }_{ Z }^{ 2 }\left( h \right)  }{ { a }_{ B }{ T }^{ 2 } } +4{ y }_{ t }^{ 4 } } \log { \frac { { m }_{ t }^{ 2 }\left( h \right)  }{ { a }_{ F }{ T }^{ 2 } }  }  \right)$}.\nonumber
\end{eqnarray}

The variance of the thermal fluctuations of the Higgs is given as~\cite{Linde:1978px, Linde:1991sk,Dine:1992wr,Dine:1992vs}
\begin{eqnarray}
\left< { h }^{ 2 } \right>_{ T} &=& \frac{1}{2\pi^{2}}\int _{ 0 }^{  \infty  }{ \frac{ { k }^{ 2 }dk }{ \sqrt {  { k }^{ 2 }+m _{ \rm eff }^{ 2 }  }\left[e^{ \frac{\sqrt { { k }^{ 2 }+m _{ \rm eff }^{ 2 } }}{T}}-1 \right]  }  }\nonumber \\
&\simeq& \frac { { T }^{ 2 } }{ 12 } -\frac { { m }_{\rm eff }T }{ 4\pi  },
\label{eq:kzbbbb}
\end{eqnarray}
where the thermal Higgs mass is $m _{ \rm eff}=c_{T}^{1/2}T$ and numerically we obtain $c_{T}\simeq0.2$.

%%%%%%%%%%%%%%%%%%%%%%%%%%%%%%%%%%%%%%%%%%%%%%%%%%%%%%%%%%%%%%%%%%%%%%
\begin{figure}[t]
\includegraphics[width=87mm]{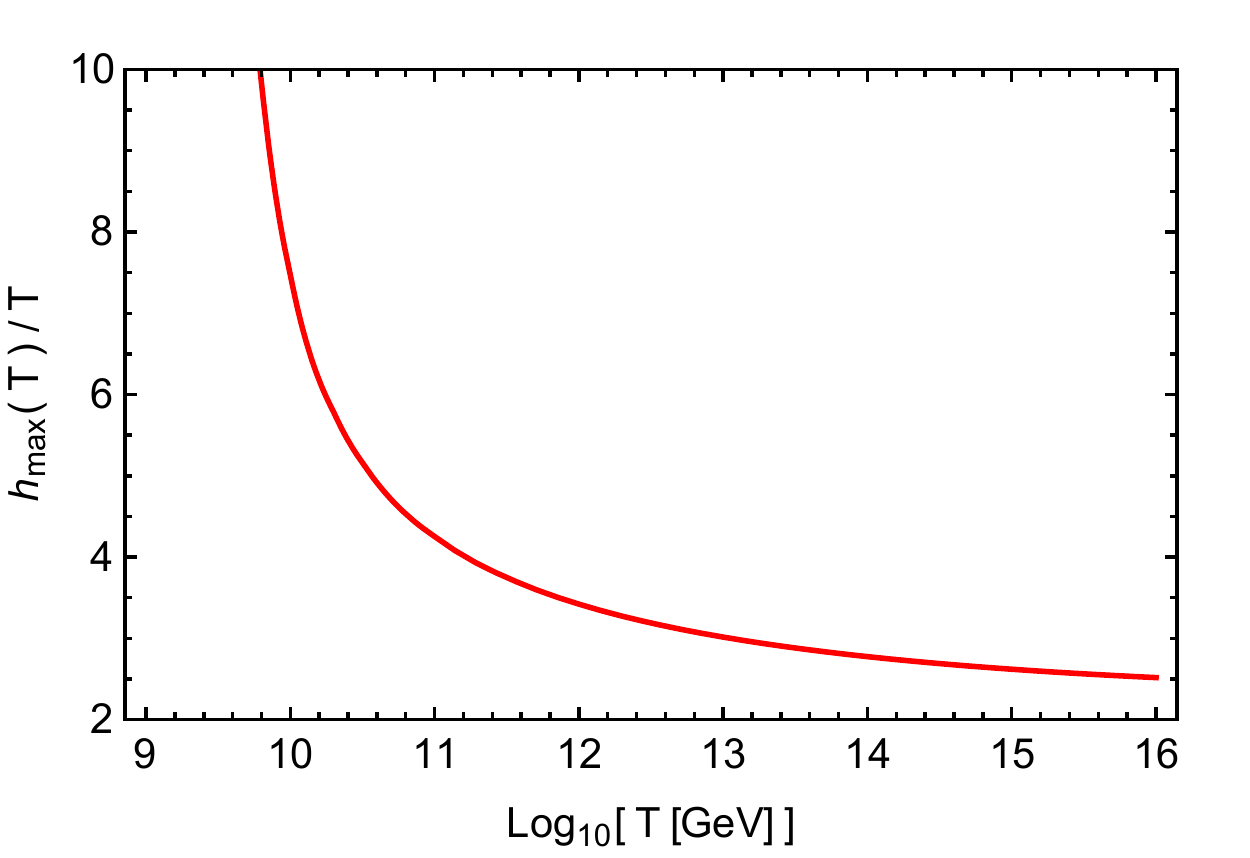}
\caption{Plot of $h_{\rm max}\left(T\right)/T$ by using RG improved effective 
Higgs potential at finite temperature. }
\label{Fig:7}
\end{figure}
%%%%%%%%%%%%%%%%%%%%%%%%%%%%%%%%%%%%%%%%%%%%%%%%%%%%%%%%%%%%%%%%%%%%%

We can estimate the relation with the thermal Higgs fluctuation and the
physical probability of the Higgs AdS domains shown in (\ref{eq:hssssdg}) as 
\begin{equation}
\frac{\left< { h }^{ 2 } \right>_{T}}{{ h }_{ \rm max }^{2}\left( { T } \right)}< \frac{1}{6N_{\rm hor}}\label{eq:tttttd}.
\end{equation}
The maximum of the Higgs potential is moved out to larger values of $h$ when thermal corrections are taken into account, and numerically we have found that $h_{\rm max}$ can be well estimated as $h_{\rm max}\left( { T } \right)= 2\sim6 \ T$. 
In Fig.~\ref{Fig:7}, we show $h_{\rm max}\left(T\right)/T$ by using the RG improved Higgs potential at the high temperature.

For simplicity, we consider the following condition 
\begin{equation}
\frac{6N_{\rm hor} \left< { h }^{ 2 } \right>_{T}}{{ h }_{ \rm max }^{2}\left( { T } \right)}< 1\label{eq:tfsgsg}.
\end{equation}
In Fig.~\ref{Fig:7}, we assume $N_{\rm hor}=60$
\footnote{$e^{3N_{\rm hor}}$ corresponds to the physical volume of our universe at the end of the inflation.}
and plot $6N_{\rm hor}
\left< { h }^{ 2 } \right>_{T}/ { h }_{ \rm max }^{2}\left( { T } \right)$
by using the RG improved effective Higgs potential at the high temperature.
When we set $N_{\rm hor}=60$, the constraint \eqref{eq:tfsgsg} gives us the 
following upper bound on the temperature $T$ 
\begin{equation}
T < 2.4\times10^{10} {\rm GeV} \label{eq:dfhjdh}.
\end{equation}

It was previously thought that thermal Higgs fluctuations do not
destabilize the standard electroweak vacuum because the probability for
the thermal vacuum decay of one Hubble-sized region via the instanton methods is sufficiently
small~\cite{Anderson:1990aa,Arnold:1991cv,Espinosa:1995se,Rose:2015lna}. 
However, after inflation, there are the large classical Higgs field and the early Universe contains 
huge number of independent Hubble-horizon regions.
Therefore, the total decay probability 
due to the thermal Higgs fluctuations would be worse.
Although, in this paper, we don't conclude whether the variance of the thermal Higgs fluctuation destabilize or not,
it is necessary to investigate thoroughly the thermal vacuum metastability during the reheating era.
We plan to perform a detailed analysis of the stochastic approach and
instanton methods in a separate paper.
In the rest of this section, we assume that the variance of thermal Higgs fluctuations
destabilize the standard electroweak vacuum and show how 
the Hubble scale $H$ is restricted in this case.

%%%%%%%%%%%%%%%%%%%%%%%%%%%%%%%%%%%%%%%%%%%%%%%%%%%%%%%%%%%%%%%%%%%%%%
\begin{figure}[t]
\includegraphics[width=87mm]{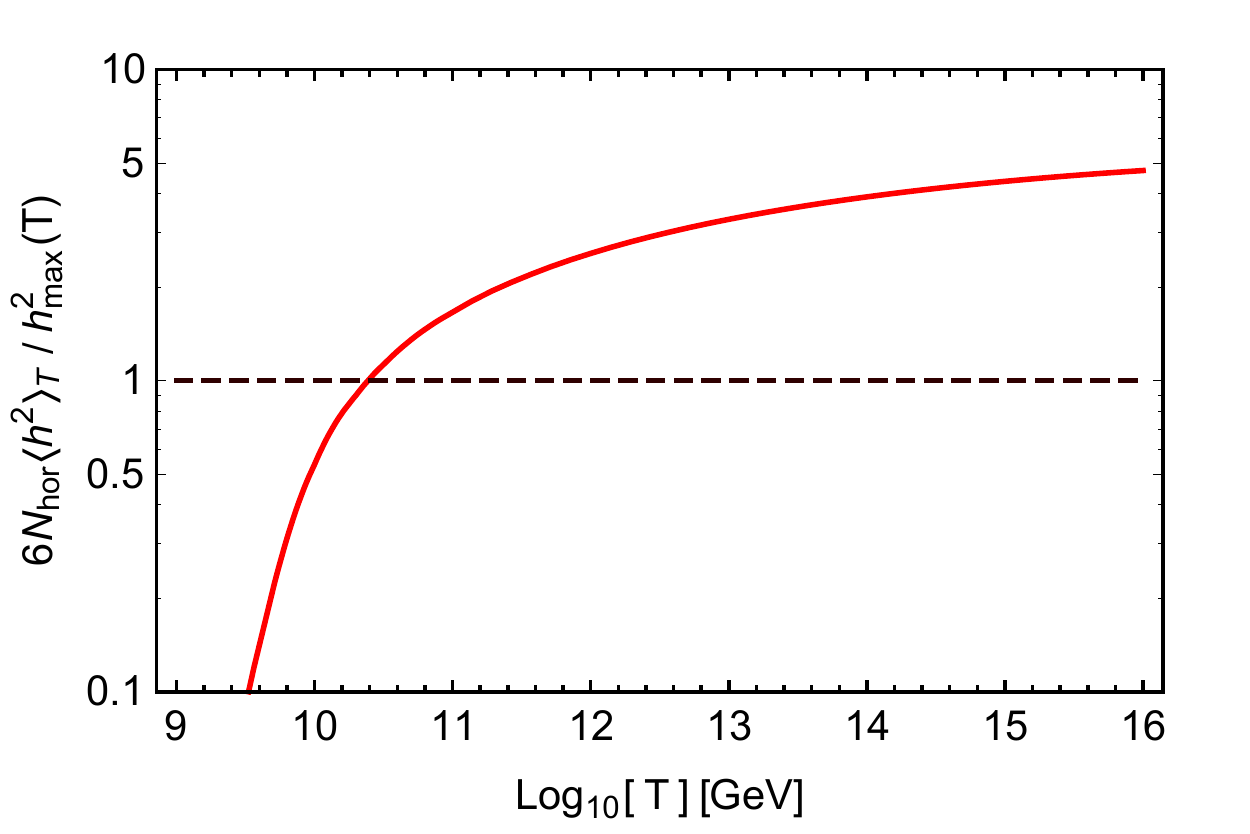}
\caption{We set $N_{\rm hor}=60$ and plot $6N_{\rm hor}
\left< { h }^{ 2 } \right>_{T}/ { h }_{ \rm max }^{2}\left( { T } \right)$
by using the RG improved Higgs potential at the high temperature.
We obtain the constraint of the temperature $T < 2.4\times10^{10} {\rm GeV}$.}
\label{Fig:8}
\end{figure}
%%%%%%%%%%%%%%%%%%%%%%%%%%%%%%%%%%%%%%%%%%%%%%%%%%%%%%%%%%%%%%%%%%%%%%

It is known that the reheating temperature $T_{\rm reh}$ is not the
maximal temperature, unless the reheating process is
instantaneous. Just after inflation, although still sub-dominant, the
decay products from the oscillating inflaton field can become thermalized
and produce a so-called dilute plasma. Then, the maximal temperature
$T_{\rm max}$ can be estimated
by~\cite{Chung:1998rq,Giudice:2000ex,Kolb:2003ke}
\begin{equation}
\scalebox{0.94}{$\displaystyle { T }_{\rm max }=\left( \frac { 3 }{ 8 }  \right)^{2/5} \left( \frac { 40 }{ { \pi  }^{ 2 } }  \right)^{1/8} \frac { { g }_{ * }^{ 1/8 }\left( { T }_{ \rm reh } \right)  }{ { g }_{ * }^{ 1/4 }\left( { T }_{ \rm max } \right)  } { M }_{ \rm pl }^{ 1/4 }H_{\rm end }^{ 1/4 }{ T }_{ \rm reh }^{ 1/2 }$},\label{eq:erherttt}
\end{equation}
with the reduced Planck mass $M_{\rm pl}=2.4\times10^{18}\ {\rm GeV}$
and $g_{*}\left( { T } \right)$ is the number of relativistic degrees of
freedom at the temperature $T$. 
By using constraints \eqref{eq:dfhjdh} and assuming $T_{\rm reh} < T_{\rm max} $, we can
obtain the upper bound on the Hubble scale $H$ as a function of $T_{\rm reh} $.

%%%%%%%%%%%%%%%%%%%%%%%%%%%%%%%%%%%%%%%%%%%%%%%%%%%%%%%%%%%%%%%%%%%%%%
\section{Conclusion}
%%%%%%%%%%%%%%%%%%%%%%%%%%%%%%%%%%%%%%%%%%%%%%%%%%%%%%%%%%%%%%%%%%%%%%
In this paper, we have discussed the stability of the Higgs vacuum
during primordial inflation, preheating, and reheating.  In the absence
of any corrections to the Higgs potential, inflationary vacuum
fluctuations of the Higgs field can easily destabilize the standard
electroweak vacuum and produce a lot of AdS domains. If a relatively
large nonminimal Higgs-gravity coupling or inflaton-Higgs coupling is
introduced, a sizable effective mass term is induced, which raises the
effective Higgs potential and weakens the Higgs field
fluctuations. Therefore, it is possible to suppress the formation of
Higgs AdS domains during inflation.  However, after inflation, such
effective masses are ineffective for stabilizing the large Higgs field.
Moreover, nonminimal Higgs-gravity coupling
and inflaton-Higgs coupling can also give rise to the generation of
large Higgs fluctuations after inflaton via parametric resonance. Hence,
such couplings cannot suppress the formation of Higgs AdS
domains. We find that the parametric resonance during preheating 
excludes values of the nonminimal coupling and inflaton-Higgs coupling
as $\xi <10^{1.6}$ and $g <10^{-4}$.
Furthermore, thermal Higgs fluctuations during the reheating stage 
cannot be neglected on the electroweak vacuum metastability.  
Our results show that the thermal Higgs fluctuations produce AdS domains 
in the reheating stage unless $T < 2.4\times10^{10} {\rm GeV}$.
We conclude that through the epochs of inflation, preheating and reheating,
a lot of Higgs AdS domains are inevitably produced unless the energy scale of the inflaton potential
is much smaller than the GUT scale, or the effective Higgs potential is stabilized below the Planck scale.

%%%%%%%%%%%%%%%%%%%%%%%%%%%%%%%%%%%%%%%%%%%%%%%%%%%%%%%%%%%%%%%%%%%%%%
\acknowledgments
%%%%%%%%%%%%%%%%%%%%%%%%%%%%%%%%%%%%%%%%%%%%%%%%%%%%%%%%%%%%%%%%%%%%%%
We would like to thank Satoshi Iso, Kyohei Mukaida, Kazunori Nakayama, Mihoko M. Nojiri, Kengo Shimada and Jonathan White.
This work is supported in part by MEXT KAKENHI No.15H05889 and No.16H00877 (K.K.), and JSPS KAKENHI
Nos.26105520 and 26247042 (K.K.). The work of K.K. is also supported
by the Center for the Promotion of Integrated Science (CPIS) of
Sokendai (1HB5804100).

%%%%%%%%%%%%%%%%%%%%%%%%%%%%%%%%%%%%%%%%%%%%%%%%%%%%%%%%%%%%%%%%%%%%%%
\appendix
%%%%%%%%%%%%%%%%%%%%%%%%%%%%%%%%%%%%%%%%%%%%%%%%%%%%%%%%%%%%%%%%%%%%%%

%%%%%%%%%%%%%%%%%%%%%%%%%%%%%%%%%%%%%%%%%%%%%%%%%%%%%%%%%%%%%%%%%%%%%%
\section{RG improved effective potential}
%%%%%%%%%%%%%%%%%%%%%%%%%%%%%%%%%%%%%%%%%%%%%%%%%%%%%%%%%%%%%%%%%%%%%%
In this appendix we provide the RG-improved effective
potential for the Higgs~\cite{Ford:1992mv,Casas:1994qy,Espinosa:1995se}, which 
is written in the $\rm \overline {MS} $ scheme and in the 't Hooft-Landau gauge as
\begin{equation}
V_{\rm eff}\left( h \right)=V_{\rm tree}\left( h \right)+V_{\rm 1-loop}\left( h \right).
\end{equation}
The improved tree-level correction to the effective Higgs potential take the form,
\begin{equation}
V_{\rm tree}\left( h \right)=\frac{1}{4}\lambda\left( t \right)h^{4}\left( t \right),
\end{equation}
where the running Higgs field is $h\left( t \right)=G\left( t \right)h$.
The wavefunction renormalization factor $G\left( t \right)$ is
given in terms of the anomalous dimension $\gamma$ as
\begin{equation}
G\left( t \right)=\exp \left( -\int _{ 0 }^{ t }{ \gamma \left( t' \right) dt' }  \right) ,
\end{equation}

The one-loop correction to the effective Higgs potential at zero-temperature is 
\begin{equation}
V_{\rm 1-loop}\left( h \right)=\sum _{ i=W,Z,t }^{  }{ \frac { { n }_{ i } }{ 64{ \pi  }^{ 2 } } { m }_{ i }^{ 4 }\left( h \right) \left[ \ln { \frac { { m }_{ i }^{ 2 }\left( h \right)  }{ { \mu  }^{ 2 }\left( t \right)  } -{ C }_{ i } }  \right]  },
\end{equation}
where the number of degrees of freedom $n_{i},~ i = W,Z,t$,
and the coefficients $C_{i},~ i = W,Z,t$  are given by
\begin{eqnarray}
n_{W}&=&6,\ n_{Z}=3,\ n_{t}=-12, \nonumber \\
C_{W}&=&C_{Z}=5/6,\ C_{t}=3/2.
\end{eqnarray}

The masses of $W$, $Z$ and $t$ depend on the background Higgs
field value $h$ as follows
\begin{eqnarray}
m_{W}^{2}\left( h \right)&=&\frac{g^{2}\left( t \right) }{4}h^{2}\left( t \right),\\ 
m_{Z}^{2}\left( h \right)&=&\frac{g^{2}\left( t \right)+g'^{2}\left( t \right)}{4}h^{2}\left( t \right),\\ 
m_{t}^{2}\left( h \right)&=&\frac{y_{t}^{2}\left( t \right)}{2}h^{2}\left( t \right),
\end{eqnarray}
where $g$, $g'$ and $y_{t}$ are the $SU(2)_{L}$, $U(1)_{Y}$, and top
Yukawa couplings, respectively.

We calculate the $\beta$ functions and the
anomalous dimension $\gamma$ to two-loop order in the current
study.  The $\beta$ functions for a generic coupling parameter X are defined through the relation
\begin{equation}
\frac { dX\left( t \right) }{ dt  } =\sum _{i }^{  }{{ \beta  }_{ X}^{ \left( i \right) }}.
\end{equation}
The $\beta$ functions and anomalous dimension $\gamma$ at
one- and two-loop order are given as follows~\cite{PhysRevD.46.5206,MACHACEK198570,Machacek1984221,Machacek198383,Ford:1992pn,Hertzberg:2012zc,Holthausen:2011aa}:
{\allowdisplaybreaks[1]
\begin{widetext}
\begin{eqnarray}
 { \beta  }_{ \lambda  }^{ \left( 1 \right)  }&=&\frac { 1 }{ { \left( 4\pi  \right)  }^{ 2 } }\left[ \lambda \left( -9{ g }^{ 2 }-3{ g' }^{ 2 }+12{ y }_{ t }^{ 2 } \right) +24{ \lambda  }^{ 2 }+\frac { 3 }{ 4 } { g }^{ 4 }+\frac { 3 }{ 8 } { \left( { g }^{ 2 }+{ g' }^{ 2 } \right)  }^{ 2 }-6{ y }_{ t }^{ 4 }\right],\\
{ \beta  }_{ { y }_{ t } }^{ \left( 1 \right)  }&=&\frac { 1 }{ { \left( 4\pi  \right)  }^{ 2 } }\left[ \frac { 9 }{ 2 } { y }_{ t }^{ 3 }+{ y }_{ t }\left( -\frac { 9 }{ 4 } { g }^{ 2 }-\frac { 17 }{ 12 } { g' }^{ 2 }-8{ g }_{ s }^{ 2 } \right)\right],\\
{ \beta  }_{ { g } }^{ \left( 1 \right)  }&=&\frac { 1 }{ { \left( 4\pi  \right)  }^{ 2 } }\left[-\frac { 19 }{ 6 } { g }^{ 3 }\right],\quad{ \beta  }_{ { g' } }^{ \left( 1 \right)  }=\frac { 1 }{ { \left( 4\pi  \right)  }^{ 2 } }\left[\frac { 41 }{ 6 } { g' }^{ 3 }\right],
\quad{ \beta  }_{ { g }_{ s } }^{ \left( 1 \right)  }=\frac { 1 }{ { \left( 4\pi  \right)  }^{ 2 } }\biggl[-7{ g }_{ s }^{ 3 }\biggl],\\
{ \beta  }_{ \lambda  }^{ \left( 2 \right)  }&=&\frac { 1 }{ { \left( 4\pi  \right)  }^{ 4} }\biggl[-312{ \lambda  }^{ 3 }-144{ \lambda  }^{ 2 }{ y }_{ t }^{ 2 }+36{ \lambda  }^{ 2 }\left( { 3g }^{ 2 }+{ g' }^{ 2 } \right) -3\lambda { y }_{ t }^{ 4 }+\lambda { y }_{ t }^{ 2 }\left( \frac { 45 }{ 2 } { g }^{ 2 }+\frac { 85 }{ 6 } { g' }^{ 2 }+80{ g }_{ s }^{ 2 } \right)\nonumber \\&& -\frac { 73 }{ 8 } \lambda { g }^{ 4 }+\frac { 39 }{ 4 } \lambda { g }^{ 2 }{ g' }^{ 2 }+\frac { 629 }{ 24 } \lambda { g' }^{ 4 }+30{ y }_{ t }^{ 6 }-32{ y }_{ t }^{ 4 }{ g }_{ s }^{ 2 }-\frac { 9 }{ 4 } { y }_{ t }^{ 2 }{ g }^{ 4 }-\frac { 8 }{ 3 } { y }_{ t }^{ 4 }{ g' }^{ 2 }\nonumber \\&&+\frac { 21 }{ 2 } { y }_{ t }^{ 2 }{ g }^{ 2 }{ g' }^{ 2 }-\frac { 19 }{ 4 } { y }_{ t }^{ 2 }{ g' }^{ 4 }+\frac { 305 }{ 16 } { g }^{ 6 }-\frac { 289 }{ 48 } { g }^{ 4 }{ g' }^{ 2 }-\frac { 559 }{ 48 } { g }^{ 2 }{ g' }^{ 4 }-\frac { 379 }{ 48 } { g' }^{ 6 }\biggl],\\ 
{ \beta  }_{ { y }_{ t } }^{ \left( 2 \right)  }&=&\frac { 1 }{ { \left( 4\pi  \right)  }^{ 4 } }\biggl[{ y }_{ t }\Bigl( -12{ y }_{ t }^{ 4 }+{ y }_{ t }^{ 2 }\left( \frac { 225 }{ 16 } { g }^{ 2 }+\frac { 131 }{ 16 } { g' }^{ 2 }+36{ g }_{ s }^{ 2 }-12\lambda  \right) +\frac { 1187 }{ 216 } { g' }^{ 4 }\nonumber \\&&-\frac { 3 }{ 4 } { g }^{ 2 }{ g' }^{ 2 }+\frac { 19 }{ 9 } { g' }^{ 2 }{ g }_{ s }^{ 2 }-\frac { 23 }{ 4 } { g }^{ 4 }+9g^{ 2 }{ g }_{ s }^{ 2 }-108{ g }_{ s }^{ 4 }+6{ \lambda  }^{ 2 }\Bigl)\biggl],\\ 
{ \beta  }_{ g }^{ \left( 2 \right)  }&=&\frac { 1 }{ { \left( 4\pi  \right)  }^{ 4 } }\left[ { g }^{ 3 }\left( \frac { 35 }{ 6 } { g }^{ 2 }+\frac { 3 }{ 2 } { g' }^{ 2 }+12{ g }_{ s }^{ 2 }-\frac { 3 }{ 2 } { y }_{ t }^{ 2 } \right)\right],\\
{ \beta  }_{ g' }^{ \left( 2 \right)  }&=&\frac { 1 }{ { \left( 4\pi  \right)  }^{ 4 } }\left[{ g' }^{ 3 }\left( \frac { 9 }{ 2 } { g }^{ 2 }+\frac { 199 }{ 18 } { g' }^{ 2 }+\frac { 44 }{ 3 } { g }_{ s }^{ 2 }-\frac { 17 }{ 6 } { y }_{ t }^{ 2 } \right)\right],\\
{ \beta  }_{ { g }_{ s } }^{ \left( 2 \right)  }&=&\frac { 1 }{ { \left( 4\pi  \right)  }^{ 4 } }\left[{ { g }_{ s } }^{ 3 }\left( \frac { 9 }{ 2 } { g }^{ 2 }+\frac { 11 }{ 6 } { g' }^{ 2 }-26{ g }_{ s }^{ 2 }-2{ y }_{ t }^{ 2 } \right)\right], \\
{ \gamma  }^{ \left( 1 \right)  }&=&\frac { 1 }{ { \left( 4\pi  \right)  }^{ 2 } }\left[ 3{ y }_{ t }^{ 2 }-\frac { 9{ g }^{ 2 } }{ 4 } -\frac { 3{ g' }^{ 2 } }{ 4 } \right],\\ 
{ \gamma  }^{ \left( 2 \right)  }&=&\frac { 1 }{ { \left( 4\pi  \right)  }^{ 4 } }\left[6{ \lambda  }^{ 2 }-\frac { 27 }{ 4 } { y }_{ t }^{ 4 }+\frac { 5 }{ 2 } \left( \frac { 9 }{ 4 } { g }^{ 2 }+\frac { 17 }{ 12 } { g' }^{ 2 }+8{ g }_{ s }^{ 2 } \right) { y }_{ t }^{ 2 }-\frac { 271 }{ 32 } { g }^{ 4 }+\frac { 9 }{ 16 } { g }^{ 2 }{ g' }^{ 2 }+\frac { 431 }{ 96 } { g' }^{ 4 } \right]. 
\end{eqnarray}
\end{widetext}}

\bibliography{higgs}

\end{document}